 \documentclass[preprint2]{aastex}

\shorttitle{Djorgovski et al.}
\shortauthors{Collapsed Cores in Globular Clusters}

\begin{document}


\title{The COBE DIRBE Point Source Catalog}


\author{Beverly J. Smith}
\affil{Department of Physics, Astronomy, and Geology, East Tennessee State University,
    Box 70652, Johnson City, TN  37614}
\email{smithbj@etsu.edu}
\author{Stephan D. Price}
\affil{Air Force Research Laboratory,
Space Vehicles Directorate,
29 Randolph Road, Hanscom
AFB, MA 01731}
\email{Steve.Price@hanscom.af.mil}
\author{Rachel I. Baker}
\affil{Department of Physics and Astronomy, East Tennessee State University,
    Box 70652, Johnson City, TN  37614}
\email{zrih1@yahoo.edu}



\begin{abstract}

We present the COBE DIRBE Point Source Catalog, an all-sky
catalog containing infrared photometry 
in 10 infrared bands from 
1.25 $\mu$m and 240 $\mu$m for 11,788 of
the brightest near and mid-infrared point
sources in the sky.  
Since DIRBE had excellent temporal coverage (100 $-$ 1900
independent measurements per object during the 10 month
cryogenic mission), the Catalog also contains
information about variability at each wavelength, 
including amplitudes
of variation observed during the mission.
Since the DIRBE spatial resolution is relatively
poor (0.7$^{\circ}$), we 
have
carefully investigated the question of confusion,
and have flagged sources with infrared-bright companions
within the DIRBE beam.
In addition, we filtered the DIRBE light curves for
data points affected by companions outside of the main DIRBE
beam but within the `sky' portion of the scan.
At high Galactic latitudes ($|$b$|$ $>$ 5$^{\circ}$), 
the Catalog contains essentially all of the unconfused
sources with flux densities
greater than 90, 60, 60, 50, 90, and 165 Jy at 1.25, 2.2,
3.5, 4.9, 12, and 25 $\mu$m, respectively, 
corresponding to magnitude limits of approximately 3.1,
2.6, 1.7, 1.3, $-$1.3, and $-$3.5.
At longer wavelengths
and in the Galactic Plane, the completeness
is less certain because of the large DIRBE beam and 
possible contributions from extended emission.
The Catalog also contains the names of the sources 
in other catalogs,
their spectral types, variability types, and whether or not the sources
are known OH/IR stars.
We discuss a few remarkable objects in the Catalog, including
the extremely red object 
OH 231.8+4.2 (QX Pup),
an asymptotic giant branch star in transition to a
proto-planetary nebula, 
which has a DIRBE 25 $\mu$m amplitude of 0.29 $\pm$ 0.07 magnitudes.

\end{abstract}


\keywords{stars: AGB and post-AGB, stars: variable: Miras, astronomical databases: catalogs,
infrared: stars }

\section{Introduction}

Infrared surveys provide a window on the content and structure of the Galaxy, 
being relatively free of the interstellar extinction that compromises 
measurements at shorter wavelengths.  Since the nature of the 
brightest sources in the sky changes with the infrared wavelength and 
the majority of these objects are variable, measurements over a 
large wavelength baseline with a fairly dense sampling in time 
are needed to fully characterize the sources.  
At optical and near-infrared wavelengths ($\le$3 $\mu$m),
the spectral energy distributions
of stars are dominated by radiation
from the stellar photosphere, while in the mid-infrared ($\ge$6 $\mu$m),
emission from circumstellar dust becomes important.  
In the near-infrared (0.9 $-$ 2.2 $\mu$m), the sky
was first surveyed by the Two Micron Sky Survey
(TMSS; \citet{nl69}), and more recently
by the 
Two Micron All Sky Survey (2MASS; \citet{c03})
and the DENIS survey \citep{ep99}.
The Infrared Astronomical Satellite (IRAS) survey 
covered almost the whole sky in the 12 $-$ 100
$\mu$m spectral range 
\citep{b88}.
The small gaps in the IRAS mid-infrared coverage
have since been filled in by the 
Midcourse Space Experiment (MSX) mission \citep{p99}.
The most complete survey made so
far 
in the 
intermediate spectral region between 3 and 6 $\mu$m 
was 
the
Air Force
Geophysical Laboratory (AFGL)
Infrared Sky Survey, which covered 71$\%$
of the sky to a sensitivity of 90 Jy 
at 4.2 $\mu$m 
\citep{pw76}.
The more sensitive 
MSX survey \citep{p99}
covered only $\sim$15$\%$ of the sky 
(two-thirds of which was
within 6$^{\circ}$ of the Galactic Plane) 
to a point source limit
at 4.3 $\mu$m
of $\sim$20 Jy.

There is another untapped infrared database suitable
for the construction of an infrared point source catalog:
the archival data
from the Diffuse Infrared Background Experiment (DIRBE) \citep{h98}
on the Cosmic
Background Explorer (COBE) satellite \citep{b92}.
DIRBE operated at cryogenic temperatures
for 10 months in 1989-1990, providing full-sky
coverage at 10 infrared wavelengths
(1.25, 2.2, 3.5, 4.9, 12, 25, 60, 100, 140, and 240 $\mu$m).
Although DIRBE was designed to 
search for the cosmic infrared background, the data are also useful for
studying point sources, in spite of its relatively 
poor spatial resolution (0.7$^{\circ}$).
To date, however,
point source fluxes from DIRBE have been little
utilized.
During the mission,
DIRBE stellar fluxes were mainly used for 
calibration verification \citep{bm97, c98}.
Recently, we used the DIRBE database to extract high quality
1.25 $-$ 25 $\mu$m time-sequence 
infrared photometry for 38 known Mira variable stars \citep{smith02}
and 207 high Galactic latitude 12 $\mu$m-selected
sources \citep{smith03}, while \citet{knapp03} extracted DIRBE 2.2 $\mu$m
light curves for a few dozen known variable stars.
These studies showed that DIRBE provided good stellar photometry
in the six shortest wavelengths.
At 4.9 $\mu$m, 
the sensitivity per measurement
is $\sim$10 Jy
\citep{smith02},
about a magnitude
fainter than the AFGL survey.   
Thus DIRBE covers the 85$\%$ of the sky
not surveyed by MSX to a sensitivity per scan at $\sim$4 $\mu$m
similar to that of MSX.

Another DIRBE advantage is 
its excellent temporal
coverage, that permits a careful investigation of possible 
infrared variability
of stars.
A typical location on the sky was observed 10 $-$ 15 times during the
course of a week, and, on the average, about
200 times during the 10 month cryogenic mission
\citep{h98}.
The DIRBE coverage varied with position on the sky. 
Sources near the ecliptic
poles were observed approximately twice a day, 
producing 400 $-$ 1000 observations over the mission.
Near the ecliptic plane, sources were typically observed
roughly twice a day for 2 months, then
were inaccessible for 4 months before coming back into view.
By comparison, IRAS typically provided only two or three independent
flux measurements of a star at 12 $\mu$m 
\citep{ls92},
while MSX made up to six observations
over a four month period \citep{e99}.

In this paper, we describe the DIRBE Point Source Catalog, 
a database of infrared photometry and variability information for the
brightest sources in the infrared sky. 
In the \citet{smith02} study, we specifically targeted asymptotic
giant branch (AGB) stars
that were known to be Mira variables.  
Consequently, we neglected non-Mira AGB stars,
infrared-bright non-AGB stars,
Miras
not previously classified as Miras, 
and 
very dust-enshrouded 
stars without bright optical counterparts.
In \citet{smith03}, we studied a complete 12 $\mu$m flux-limited
sample of 207 IRAS sources at Galactic latitudes 
$|$b$|$ $>$ 5$^{\circ}$.  
This study neglected bright stars at the shorter DIRBE wavelengths
or at 25 $\mu$m,
as well as stars in the Galactic plane.
In the current study, 
we constructed an all-sky DIRBE Point Source Catalog that
extends these earlier studies to include all sources above
a uniform signal to noise selection criterion in each of the six shortest
wavelength DIRBE bands, as well as sources in the
Galactic Plane.
The present Catalog 
contains essentially all of the unconfused high Galactic latitude 
point sources detected by DIRBE 
in the six shorter wavelength filters
with
S/N per individual scan greater than 3 $-$ 9.
These levels are  
sufficient to provide
good light curves.   In the Galactic plane and at longer wavelengths
the Catalog is less complete.

\section{Sample Selection}

Unlike the IRAS and 2MASS Catalogs, the DIRBE Point Source 
Catalog was not constructed by searching the DIRBE database with
a point source template and extracting
sources based on signal to noise and confirmation criteria.
The DIRBE Catalog was constructed using a target sample list
obtained 
from other infrared catalogs.  
Since DIRBE is much less sensitive per scan than IRAS 
or 2MASS, essentially
all of the point sources with high S/N light curves in 
the DIRBE database are already contained
in IRAS, 2MASS, and/or MSX.  Thus for simplicity we used these previous
catalogs to select a sample for the DIRBE Point Source Catalog.

Our initial sample included a total of 21,335 sources; the final Catalog
contains 11,788 sources.
The initial sample was selected from
the 
\citet{PSC},
the 2MASS Point Source
Catalog \citep{c03}, and/or the MSX Point Source Catalog
Version 1.2 \citep{e99} that 
satisfied at least one
of the following criteria: a) 2MASS J 
magnitude $\le$ 4.51 (F$_{1.25}$ $\ge$ 25 Jy),
b) 2MASS K magnitude $\le$ 3.81 (F$_{2.2}$ $\ge$ 20 Jy), 
c) IRAS or MSX F$_{12}$ $\ge$ 15 Jy,
or d) IRAS or MSX F$_{25}$ $\ge$ 27.5 Jy.
The 1.25 and 2.2 $\mu$m
limits are equal to the 
average
1$\sigma$ sensitivity per scan
in the raw DIRBE light curves of \citet{smith02}, while
the 12 and 25 $\mu$m limits are
0.5 $\times$ the average noise levels per scan in that study.
These low limits were selected in order to avoid missing variable stars
that may have been faint during the 2MASS, IRAS, or MSX mission,
and to improve
the completeness at 3.5 and 4.9 $\mu$m
(see Section 6).
Since the filtering process improves the average per-measurement uncertainty
(see Section 5), a sensitive selection criterion is warranted to include
as many sources as possible.

There are 7,872 sources with 2MASS J $\le$ 4.51, 
20,492 sources with 2MASS K $\le$ 3.81, 4,969 sources with IRAS F$_{12}$ $\ge$ 15 Jy, 
40 sources in the MSX IRAS Gaps survey with MSX F$_{12}$ $\ge$ 15 Jy,
2,753 sources with IRAS F$_{25}$ $\ge$ 27.5 Jy, and 18 sources in the
MSX IRAS Gaps survey with MSX F$_{25}$ $\ge$ 27.5 Jy.
Thus our initial list is dominated by stars selected by the 
2MASS criteria.
These lists were merged together to make a single target list,
containing 21,335 sources.
To merge the 2MASS and IRAS/MSX lists, we used a 60$''$ matching radius.
If more than one 2MASS source was within 60$''$ of the IRAS
position, we assumed the brightest K band source was the match.

Note we did not include sources in our input list based on their
60 and/or 100 $\mu$m IRAS flux densities, as extended emission
from cirrus becomes more significant at these wavelengths.
This means that
the DIRBE Point Source Catalog is biased against very cold objects,
such as galaxies and molecular clouds.
Since we only used the point source catalogs of 2MASS, IRAS, and MSX
for source selection, our sample is also biased against extended objects.
Note also that we are 
only targeting sources bright enough
to detect their possible variability in the DIRBE database (i.e., sources that
may be detected in a {\it single} DIRBE scan at at least one
DIRBE wavelength).  By coadding
the full light curves, it may be possible to detect fainter objects in
the DIRBE database, but without variability information and with
a higher likelihood of confusion.
Such co-addition is beyond the scope of the current Catalog.

\section{Extraction of the DIRBE Light Curves and the Catalog }

For all 21,335 sources in our list,
we extracted light curves at all ten wavelengths
from the DIRBE Calibrated Individual Observations database.
This database contains the
individual calibrated 1/8th second samples taken in science-survey
mode during each day of the DIRBE cryogenic mission.
For all scans that pass within 0.3$^{\circ}$ of the target position,
a linear baseline is fit to the sections 
$\pm$1.35$^{\circ}$$-$2.25$^{\circ}$
from the point
of closest approach.  The point source photometry is obtained
by subtraction of this baseline and correcting for the DIRBE beam response.
The uncertainties in the point source photometry are calculated as
the root sum square of the rms noise of the baseline,
the photometric error produced by a 1$'$ error in the in-scan and cross-scan
directions,
an error due to short-term detector gain variations, and
signal-dependent detector noise.
The number of individual
measurements per light curve ranges from 99 to 1932; the average
and median number of data points per light curve are 488 and 423, respectively.
The average uncertainties per measurement
in the raw 1.25 $-$ 240 $\mu$m light curves
are 33, 38, 27, 21, 107, 256, 1567, 3207, 8100, and 4510 Jy, respectively.

The COBE DIRBE Point Source Catalog
(Table 1\footnote{Table 1 is only available
in electronic form, at the Astrophysical Journal
Supplement web site at http://www.journals.uchicago.edu/ApJ})
contains
the time-averaged DIRBE flux densities $F_{\nu}$ in the ten
DIRBE bands for all 11,785
sources in our initial list
that had a flux at minimum in the weekly-averaged light curve
in {\it any}
of the six shortest DIRBE wavelengths
greater than 3$\times$ the average noise per datapoint,
plus three additional
sources (see Section 6).
These flux densities were calculated after filtering the light
curves (see Section 5).
The name of the object in the Catalog it was originally
selected from is also given (IRAS/MSX and/or 2MASS)
in Table 1.
Table 1 also contains the
number of individual measurements $N$ available after filtering the light
curves, 
the average uncertainty per measurement
$<$$err$$>$, and
the standard deviation $\sigma$ = $\sqrt{{\Sigma}(F_i - <F>)/(N-1)}$
of the individual flux measurements $F_i$.
Table 1 also gives
the observed amplitude $\Delta$$mag$,
the uncertainty assigned to that amplitude $\sigma$($\Delta$$mag$)
(see Section 7 for definitions),
and the 
confusion flags (see Sections 4 and 5 for definitions).
The positions given in Table 1 (and the positions used
for the extraction) came from either the 2MASS catalog or
the IRAS/MSX catalog, with the 2MASS position preferentially used
if available.
No color corrections have been applied to the data in Table 1 (see
\citet{h98} and \citet{smith03}
for a discussion of color corrections).
The complete filtered light curves for variable sources will be
published in a follow-up paper.

\section{Flagging for Confusion}

For each wavelength, three different
confusion flags may be set in the DIRBE 
Point Source Catalog (see Table 2 for a summary of the flags and
Table 3 for some basic statistics on the DIRBE Catalog, including statistics
on flagging and number of sources detected above 3$\sigma$ at each wavelength).
The first confusion flag is set when a second known infrared-bright
source is located within 0.5$^{\circ}$ of the target source.
We used limits for the brightness of the companion
of 25, 20,  20, 10, 30, 55, 320, and 765 Jy for 1.25 $-$ 100 $\mu$m,
respectively.  These are 
conservative limits, equal to the average per-measurement 1$\sigma$
noise levels of the DIRBE light curves in \citet{smith02}.
At 1.25 and 2.2 $\mu$m, we used
the
2MASS database to search for companions, while
at 12 and 25 $\mu$m we used 
the IRAS Point Source Catalog and the IRAS
Small Scale Structure Catalog \citep{hw85}.
At 3.5 $\mu$m, where there is no previous all-sky
survey, we used
the Catalog of Infrared Observations  
\citep{gps00}\footnote{Available from the
VizieR service at
http://vizier.u-strasbg.fr/viz-bin/VizieR},
a compilation of all published infrared observations
available in 1997.
At 4.9 $\mu$m,
we used both the Catalog of Infrared Observations and
the synthetic all-sky 4.2 $\mu$m catalog of \citet{ep96},
a list of estimated 4.2 $\mu$m flux densities
for 177,860 stars created by extrapolation from the IRAS catalogues,
other infrared catalogues,
and optical measurements.
Objects with companions above these flux limits
were flagged at the respective wavelength
in Table 1.
The numbers of sources with companions in the DIRBE beam above the DIRBE noise
level for each wavelength are given in Table 3.
Note that this flagging may be incomplete at 3.5 and 4.9 $\mu$m,
where no all-sky catalog is available.  An additional uncertainty
in the flagging may exist because the companion
may be variable.

To help the Catalog user assess the relative contribution to the DIRBE
flux density from the companion compared to the target source,
in 
Table 4\footnote{Table 4 is only available
in electronic form, at the Astrophysical Journal
Supplement web site at http://www.journals.uchicago.edu/ApJ}
we give the 2MASS and/or IRAS/MSX photometry of the target source,
while in 
Table 5\footnote{Table 5 is only available
in electronic form, at the Astrophysical Journal
Supplement web site at http://www.journals.uchicago.edu/ApJ}
we provide the names of the companions, their 2MASS/IRAS/CIO/Egan $\&$ Price (1996)
flux densities, and the distances between 
the companions and the target sources.
In some cases the companion's flux density is small compared to that
of the target source; in other cases, the companion dominates (see
Sections 11 and 13 for examples).

Note that when two IRAS and/or two 2MASS sources in our input list
are within a single
DIRBE beam, they are both included in the DIRBE Catalog as separate
listings, each flagged for a companion.
In the DIRBE Catalog itself,
we do not deconvolve the DIRBE flux of confused sources into that of
the two sources, however, the information available in Tables 4 and 5 
may help 
the Catalog user estimate the relative importance of each
source to the DIRBE flux.
As noted above, when both an IRAS and a 2MASS
counterpart are listed in Table 1, this association is based solely
on a positional offset of $\le$1$'$.  It is possible that in some 
cases the IRAS and 2MASS sources 
are two different objects; with the DIRBE data alone, it is impossible
to discriminate between them.
Also note that an IRAS counterpart to a 2MASS
source is only included if there is an IRAS source within 1$'$ of the 2MASS
source brighter than our IRAS selection criteria,
and vice versa.  Tables 1 and 4 do not list possible associations with
IRAS or 2MASS sources fainter than our initial flux limits.

As an additional test for confusion, 
we compared the DIRBE photometry with that
of 2MASS
at 1.25 and 2.2 $\mu$m,
and with that of IRAS at 12, 25, 60, and 100 $\mu$m.
If the DIRBE flux density at minimum (after averaging in one-week
periods) is more than 3$\sigma$ 
larger than the 2MASS or IRAS
flux densities at that wavelength, a second flag is set in Table 1.
Since the DIRBE 0.7$^{\circ}$ beam is much larger than apertures used in
2MASS  (3$''$ $-$ 14$''$) and the IRAS beams ($<$8$'$), discrepant fluxes
may indicate a second source in the DIRBE beam or
the presence of extended emission around the
source.  Alternatively, it may
be due to large-amplitude or long time-scale variability. 
This second 
flag was also set if the DIRBE maximum is less than a previous measurement
minus 3$\sigma$.
In Table 3, we provide statistics on the number of sources
per wavelength
that have this second flag set.
The Catalog of Infrared Observations and the 
\citet{ep96}
catalog were excluded from this comparison because
of possible large uncertainties in the photometry, thus this flag 
was not set at 3.5 and 4.9 $\mu$m.
This flag was also not set at 140 and 240 $\mu$m, since no all-sky
catalogs are available at those wavelengths.

\section{Filtering of the Light Curves}

Another issue
is the possibility of more distant companions, outside of the main DIRBE beam,
affecting the `sky' fluxes used in the photometry extraction.
Inspection of the DIRBE data showed that, 
if a second infrared-bright star is
between about 
0.5$^{\circ}$ $-$ 2.5$^{\circ}$ of the target star
and if a scan happened to pass near that second star,
flux
from the nearby star
sometimes contributed to the `sky flux' used to calculate
the photometry for the target source, 
causing erroneous photometry with large errorbars for the targeted star. 
Fortunately, however, scans in other directions were not affected by the 
second star.
To correct for this problem,
we filtered our data to remove scans affected by nearby stars.
We searched previous 
infrared catalogs for 
objects within 3.2$^{\circ}$ of each targeted source.
At each wavelength, for each DIRBE scan for each targeted source, 
we scaled the companion's flux density
by a Gaussian with
FWHM 0.7$^{\circ}$, weighted by the minimum distance between the
scan and the companion star.
If the weighted flux density of the companion in the respective
band was greater than 
25, 20,  20, 10, 30, 55, 320, and 765 Jy at 1.25 $-$ 100 $\mu$m, respectively,
then the scan was removed
from consideration.
We also used the IRAS
Small Scale
Structure Catalog \citep{hw85} in this filtering
process, since extended sources outside the DIRBE
beam may also affect the DIRBE photometry.

The remaining measurements with large
error bars ($\ge$3 times the average uncertainty)
were likely affected by a cosmic ray hit in the `sky' portion
of the measurement.  These large deviations were removed from the light curves by
an additional filtering process.
Generally, only a few DIRBE scans per light curve
were removed by this additional
filtering processing.  
A cosmic ray hit near the star itself may not produce
a large error bar, but instead may manifest itself as a very
discrepant flux measurement with a small
error bar.  
We also removed these points from the final light curves.
No all-sky catalogs are available at the two longest DIRBE wavelengths,
thus these light curves were just filtered for measurements with
large errors or very discordant photometry.

In some cases, filtering dramatically improved the DIRBE light curves,
removing discrepant data points and those with large errorbars.
Some example DIRBE light curves before and after filtering are shown
in Figure 1 and in \citet{smith03}.
The filtered light curves show much less scatter than the originals.
In the case of the 1.25 $\mu$m light curve of IK Tau (Figure 1),
the datapoints that were removed by filtering systematically had lower apparent
flux densities than those at similar times that were not filtered.
The filtering routine found six stars near IK Tau with
2MASS F$_{1.25 {\mu}m}$ between 55
and 62 Jy, similar to the deviations seen in the datapoints that were removed.
In the case of the 2.2 $\mu$m light curve for CW Leo (Figure 1), some very discrepant points
with large error bars have been removed.
These were caused by
R Leo, which is 3$'$ away from CW Leo.
R Leo has 
2MASS F$_{2.2 {\mu}m}$ = 5505 Jy, similar to the deviations seen in the unfiltered
light curve.
In the case of the 12 $\mu$m light curve of R Leo,
the discrepant points were caused by CW Leo, which is much brighter than R Leo
at 12 $\mu$m, varying between $\sim$20,000 Jy and $\sim$ 40,000 Jy.
Note that the bad scans occur at the same time for R Leo and CW Leo, as expected.
The scans that were not removed by filtering had position angles such that they did not pass
directly through both R Leo and CW Leo.

The filtering process somewhat reduces the average
noise level in the light curves.  Over the entire Catalog,
the average noise levels in the filtered light curves are
30, 7, 17, 7, 18, 32, 118, and 468 Jy
at 1.25 $\mu$m
to 100 $\mu$m, respectively.

Because of incompleteness and photometric uncertainties in the comparison
catalogs, and the fact that some of the nearby stars may
themselves be variable,
the flagging and 
filtering routines are not always perfect.
To test for possible residual effects from companions in the final 
filtered light
curves, 
we compared the average photometric uncertainty $<$$err$$>$
with the rms $\sigma$ for two-week intervals in the light curves.  
Since the majority of variable stars in our sample are
expected to be long-period variables (see \citet{smith03}), small time-scale
variations in the light curves are probably due to a nearby star affecting
some scans more than others.
If the $\sigma$/$<$$err$$>$ was
greater than 3 for any two week period containing at least 10
measurements, a third confusion flag was set in Table 1.
The number of sources with this flag set at each wavelength
is given in Table 3.
Note that the wavelength with the most sources with this flag set
is 3.5 $\mu$m, because of the lack of an all-sky catalog for filtering
purposes.

If a large fraction of the photometric values in a light curve were
removed by filtering, the source may be confused.  If the light curve
of a source has fewer 
than about 50 $-$ 100 measurements in 
its light curve after filtering, then the photometry
and variability parameters in the DIRBE Catalog should be considered somewhat suspect.
In some cases, if a source is very confused, all of the 
scans at a wavelength were 
deemed affected by nearby companions, and so were filtered out.
In these cases, no detection is recorded at that wavelength (specifically,
the flux density is set to $-$99.9).  

\section{Completeness of the DIRBE Catalog Input List and the Catalog Itself}

Our procedure of selecting sources 
from the IRAS, 2MASS, and MSX databases
ensures that the input list for the DIRBE Catalog
should be as complete as these catalogs for bright objects
at high Galactic latitudes.
At $|$b$|$ $>$ 5$^{\circ}$,
we estimate completeness limits of approximately
90, 60, 90, and 165 Jy at
1.25, 2.2, 12, and 25 $\mu$m, 
respectively, corresponding to magnitude limits of $\sim$3.1, 2.6, $-$1.3, 
and $-$3.5.
These are 3.3$\times$, 8.6$\times$, 5.0$\times$, and 5.1$\times$ the average
noise level per scan in the filtered light curves, and are
3.6$\times$ our 1.25 $\mu$m 2MASS selection criterion, 3.0$\times$ our
selection criterion at 2.2 $\mu$m, and 6$\times$ our selection criterion
at 12 and 25 $\mu$m.
This means that all sources with amplitudes 
$\le$1.4 magnitudes at 1.25 $\mu$m, 
$\le$1.2 magnitudes at 2.2 $\mu$m, and $\le$1.9 magnitudes at 12 and 25 $\mu$m
are included
in our sample, even if they were observed at minimum by 2MASS or
IRAS/MSX and at maximum by DIRBE.   
These amplitude limits are larger than typical values for Miras 
(see Section 12 and Table 17).

The DIRBE Catalog input list should also be complete at high Galactic
latitudes at
3.5 $\mu$m to about 60 Jy,
3.5 $\times$ the average noise level in the filtered light curves.
The reddest stars in K $-$ L (i.e., F$_{2.2}$/F$_{3.5}$)
in the \citet{smith03} sample
are the Mira carbons, with F$_{2.2}$/F$_{3.5}$ $\sim$ 0.45, 
F$_{3.5}$/F$_{4.9}$ $\sim$ 0.61, and 
F$_{4.9}$/F$_{12}$ $\sim$ 0.88.
If these stars have F$_{3.5}$ = 60 Jy 
they will have F$_{2.2}$ $\sim$ 27 Jy (K $\sim$ 3.47), and thus would be included
in the sample based on the 2.2 $\mu$m criterion (K $\le$ 3.81).
Carbon Miras fainter than $\sim$44 Jy at 3.5 $\mu$m will not meet the 2.2 $\mu$m selection
criterion, however, they
are expected to be selected by the 12 $\mu$m criterion
down to 
F$_{3.5}$ = 10 Jy.
Thus the DIRBE catalog input list should be complete at high Galactic latitudes at 3.5 $\mu$m
to at least 60 Jy 
for sources with amplitudes 
$\le$ 1.9 magnitudes.  All of the sources in the DIRBE Catalog have smaller 3.5 $\mu$m amplitudes
than this limit (see Section 12).

At 4.9 $\mu$m, we estimate a high latitude completeness limit of 50 Jy, 5$\times$
the average noise level at this wavelength.
A carbon Mira with 
F$_{4.9}$ = 50 Jy 
has an expected
F$_{12}$ = 57 Jy, well above our IRAS 12 $\mu$m selection cut-off of 15 Jy.
Thus all sources with 4.9 $\mu$m amplitudes of $\le$1.4 magnitudes would be
included in our sample to a limit of 50 Jy.

To confirm these estimates of the completeness limits of our input list,
we performed three tests.
First, to search for missing sources,
we cross-correlated our input list of 21,335 sources with the
4.3 $\mu$m sources in the 
MSX IRAS Gaps survey (the 4$\%$ of the sky missed by IRAS with
$|$b$|$ $\ge$ 6$^{\circ}$; \citet{e99}),
the 
synthetic 4.2 $\mu$m catalog of
\citet{ep96},
and 
the Catalog of Infrared Observations
at 3.5 and 4.2 $-$ 4.9 $\mu$m.
There were no sources in the MSX IRAS Gaps survey with 
MSX band B1 (4.22 $-$ 4.36 $\mu$m) or band B2 (4.24 $-$ 4.45 $\mu$m) 
flux densities
greater than our nominal completeness limit of 50 Jy which were not in our
DIRBE input list, and only 8 with flux densities $\ge$ 30 Jy.
All of these sources had very low quality detections in MSX 
band B1 (quality flag = 1;
see \citet{e99}) with 
F$_{B1}$ $\le$ 41 Jy,
and were not detected in MSX band B2.
We extracted
DIRBE photometry for these 8 sources.
None of them were detected at any wavelength by DIRBE.

Only 16 
sources in the synthetic \citet{ep96} 
catalog have estimated 4.2 $\mu$m flux densities
greater than 50 Jy, $|$b$|$ $\ge$ 5$^{\circ}$, and are not in our
initial source list for the DIRBE Catalog.
We extracted
DIRBE photometry for these 16 sources.
None were detected by DIRBE at 4.9~$\mu$m.

At 3.5 $\mu$m, 
all but four of the high latitude ($|$b$|$ $\ge$ 5$^{\circ}$) sources in the Catalog of Infrared Observations
with F$_{3.5}$ greater than our nominal completeness limit of 60 Jy
are included in our sample.
At 
4.9 $\mu$m, 
the DIRBE catalog input list does not include 181
of the 
4.2 $-$ 4.9 $\mu$m high latitude sources 
in the 
Catalog of Infrared Observations brighter than
50 Jy.
Nearly all of these apparent `missing' sources are low S/N unconfirmed detections in the AFGL
or TMSS, and thus may be spurious or extended sources, or have large positional
errors.
We extracted the DIRBE light curves for these positions, and found only
one unconfused
detection above 50 Jy at 4.9 $\mu$m.
This source, the very bright star $\theta$ Cnc, was missed 
by 2MASS because it was strongly saturated.
This star was also detected by DIRBE above our nominal
completeness limits at 1.25, 2.2, and 3.5 $\mu$m.
Another source, the semi-regular variable V1888 Cyg, was only detected
at 4.9 $\mu$m by DIRBE, with a time-averaged flux density of
34.6 Jy, below our nominal Catalog completeness limit.
These two sources were added to the DIRBE Point Source Catalog.

We also added RAFGL 2688 (the Egg Nebula) to the Catalog.  This very
dusty source
lies in the small part of the sky not observed 
by either IRAS or MSX, but was strongly detected by DIRBE
(see Section 14 and Figure 9b).
These three additions bring the number of sources in the Catalog up 
to 11,788.

We note that 
infrared-bright transient objects would be missed 
by our sample selection criteria.
For example,
the Catalog of Infrared Observations contains 
three novae
with 3.5 $\mu$m flux densities greater than our nominal completeness limit of
60 Jy, and two with 4.2 $\mu$m
flux densities greater than our nominal 4.9 $\mu$m 
completeness limit of 30 Jy.
In addition, SN1987A was detected at 4.2 $\mu$m
above 30 Jy in the months
following its appearance \citep{bouchet89},
but had faded significantly by the time
of the DIRBE mission \citep{wooden93},
and was not detected by DIRBE.

As a second test of the completeness of the DIRBE Catalog input list, 
for a selected region of the sky we extracted DIRBE
light curves for a much more sensitive sample of 2382 2MASS sources,
covering $\sim$2$\%$ of the sky.
This deeper sample had a 
2MASS K band limit of F$_{2.2}$ = 1.4 Jy (K $\le$ 6.7) 
(0.02 $\times$ the average noise level in the filtered DIRBE light curves) or
F$_{1.25}$ = 12 Jy (J $\le$ 5.3) (0.4 $\times$ the DIRBE per-point uncertainty),
R.A. $\le$ 0$^{\rm h}$ 30$^{\rm m}$, $|b|$ $\ge$ 5$^{\circ}$, and all
Declinations.
None of these additional sources have unconfused DIRBE time-averaged flux densities
greater than our nominal completeness limits.
This indicates that our initial selection criteria
provides a very complete sample above these limits.

To further investigate the completeness levels of the Catalog,
in Figure 2 
we have plotted log N$-$log S at each wavelength 
for high Galactic latitudes ($|$b$|$ $\ge$ 5$^{\circ}$).
At 1.25 $-$ 25 $\mu$m, clear turn-overs are visible in these plots
at flux levels of $\sim$40, 40, 40, 25, 65, and 110 Jy, respectively.
These are somewhat lower than our nominal completeness limits, 
showing that our completeness estimates are reasonable.
The log N$-$log S plots for 60 $-$ 240 $\mu$m do not show
turn-overs, showing that, as expected, 
the Catalog is not complete at those wavelengths.

Our estimated completeness limits only pertain to unconfused sources; as noted
earlier, in some cases 
all of the measurements in a light curve are filtered out (see Section
5 and Table 6).  Since filtering does not depend upon the brightness
of the target source, the lack of photometry for these sources in the Catalog will
lower the log N$-$log S curve, but should not strongly affect the turn-over
flux density.
To investigate the incompleteness of the Catalog itself due to this
filtering, we searched
the DIRBE Catalog input list
for sources with 2MASS or IRAS/MSX photometry brighter
than our nominal completeness limit
at a given wavelength, which had all their measurements at that
wavelength removed by
filtering.  In Table 6, we provide 
statistics on the resulting incompleteness of the
Catalog.
Approximately one quarter of the photometry is lost
at 1.25, 2.2 $\mu$m, and 12 $\mu$m due to filtering,
about half at 25 $\mu$m,
about three quarters at 60 $\mu$m, and $\sim$85$\%$ at 100 $\mu$m.

This removal of all measurements by filtering 
is a strong function
of Galactic latitude, in that it is more likely
to happen at lower latitudes.
In Figure 3, using
equal sky-area bins, we plot, as function
of Galactic latitude, the fraction of sources in our input list
with 2MASS/IRAS/MSX fluxes above our nominal completeness limits that
have had all their measurements
removed by filtering.
The worst cases are for 1.25 $\mu$m and 2.2 $\mu$m, where $\sim$65$\%$
of the photometry is lost at $|$b$|$ $\le$ 5$^{\circ}$, decreasing
to less than 10$\%$ at $|$b$|$ $>$ 30$^{\circ}$.

The resulting incompleteness of the Catalog at 1.25 $\mu$m and 2.2 $\mu$m is also visible
in Figure 4, where we plot a histogram of the number of sources in
the Catalog as a function of Galactic latitude in equal sky-area bins.
There are clear turn-overs of the 1.25 $\mu$m and 2.2 $\mu$m
source counts at $|$b$|$ $\sim$ 20$^{\circ}$, showing incompleteness below
these latitudes.

In Figure 4, we also plot
the number of flagged sources as a function of latitude, while
in Figure 5,
we give the fraction of Catalog sources that are flagged 
as a function of Galactic latitude.
At 1.25 $-$ 4.9 $\mu$m, this fraction decreases with increasing latitude.
The most extreme case is that of 2.2 $\mu$m, where 95$\%$ of sources with
$|$b$|$ $<$ 5$^{\circ}$ are flagged, and $\sim$40$\%$ at $|$b$|$ $\ge$ 45$^{\circ}$.
At 12 and 25 $\mu$m, the percentage of flagged sources is highest at $|$b$|$ $\le$ 5$^{\circ}$,
but there is little correlation with latitude otherwise.

\section{The DIRBE Variability Parameters }

As noted in Section 3,
in addition to the time-averaged infrared flux densities,
Table 1 also
includes both the standard deviation $\sigma$ of the individual flux values in the
light curve of the object (after filtering) and
the mean uncertainty $<$$err$$>$ of the individual
data points in the light curve.  The comparison of these two values
provides an estimate of the likelihood
of variability of the object.
In addition, 
for each light curve with minimum flux density greater than
three times the average uncertainty per measurement,
Table 1 also includes the 
total
{\it observed} amplitude of variation during
the DIRBE observations 
$\Delta$$mag$ = $2.5 log (F_{max}/F_{min})$,
where $F_{min}$ and $F_{max}$ are the minimum and maximum flux densities
after averaging over
one week time intervals.
Table 1 also lists 
$\sigma$($\Delta$$mag$), 
the uncertainty on 
$\Delta$$mag$, 
calculated from
$\sigma$($\Delta$$mag$) = $2.5 {\sigma}_{F_{max}/F_{min}}/(2.303 (F_{max}/F_{min}))$,
where ${\sigma}_{F_{max}/F_{min}}/(F_{max}/F_{min})$ = 
$\sqrt{({\sigma}_{F_{max}}/F_{max})^2
+({\sigma}_{F_{min}}/F_{min})^2}$ and $\sigma_{F_{max}}$ and $\sigma_{F_{min}}$ are set
equal to the average uncertainty per measurement $<$$err$$>$.
If the S/N at minimum in the weekly-averaged light curve 
was less than 3 then an amplitude and amplitude uncertainty were not calculated 
(that is, they were set to $-$99.9 in Table 1).
We note that the observed changes in brightness given in the DIRBE
Catalog may not 
represent
the full range of variation for these stars,
because many of the light curves are not complete and may not cover
a full pulsation period.  In these cases, the observed variations
are a lower limit to the true amplitudes of variation.
In Table 3, we list the numbers of unflagged sources 
at each wavelength
with at least 100 measurements in their filtered light curves and
$\Delta$(mag)/$\sigma$($\Delta$mag) $\ge$ 3.

The largest unconfused DIRBE amplitudes are 2.2 magnitudes at 1.25 $\mu$m
for the oxygen-rich Mira star IK Tau (see Figure 1), 1.8 and 1.6 magnitudes at 2.2 and 3.5 $\mu$m,
respectively, for 
the carbon star CW Leo (see Figure 1), 
1.5 magnitudes at 4.9 $\mu$m for 
the 
Mira star IZ Peg (see Figure 6), 
and 
1.2 and 1.0 magnitudes at 12 and 25 $\mu$m, respectively, for
the OH/IR star OH 348.2-19.7 (see \citet{smith03}).

\section{DIRBE vs. 2MASS Photometry}

Stars brighter than 
J $\sim$ 4.5 (F$_{1.25 {\mu}m}$ $\sim$ 25 Jy)
and
K $\sim$ 3.5 (F$_{2.2 {\mu}m}$ $\sim$ 26 Jy) were saturated in 2MASS,
thus their 2MASS photometry is somewhat uncertain
($\sigma$ $\sim$ 0.2 $-$ 0.3 magnitudes;
\citet{c03}).
This means that for the brightest stars in the sky, DIRBE provides more accurate
photometry at these wavelengths for unconfused sources.
In Figure 7, 
we plot the distribution of 
the 
2MASS and DIRBE 
1.25 and 2.2 $\mu$m uncertainties
for the subset of 791 sources in the DIRBE Catalog with
2MASS K $<$ 1.0 (i.e., F$_{2.2}$ $>$  266 Jy).
There is a dramatic difference, with the median 1.25 $\mu$m uncertainty
in DIRBE being 0.03 magnitudes, and the median in 2MASS being 0.26 magnitudes.
At 2.2 $\mu$m, the median DIRBE uncertainty is 0.02 magnitudes, and the
median 2MASS uncertainty is 0.23 magnitudes.
Thus for the brightest near-infrared sources, the DIRBE photometry is
$\sim$10 times more precise than that of 2MASS.

In Figure 8a, we plot the 2MASS J magnitudes against the DIRBE 1.25 $\mu$m magnitudes
for all unflagged $|$b$|$ $\ge$ 5$^{\circ}$ 
sources in the DIRBE Catalog.
In Figure 8b, we compare the 2MASS K magnitudes with the unflagged DIRBE photometry.
For the unflagged sources in the K $<$ 1 subset, the
best-fit relationships between the 2MASS and DIRBE photometry
are J$_{DIRBE}$ = (1.011 $\pm$ 0.010)J$_{2MASS}$ $-$ 0.019 $\pm$ 0.016 
($\chi$$^2$ = 220
with $N$ = 373)
and K$_{DIRBE}$ = (1.010 $\pm$ 0.012)K$_{2MASS}$ + 0.085 $\pm$ 0.011
($\chi$$^2$ = 154
with $N$ = 389).
For the full Catalog, the best-fit relationships for the unflagged sources 
are J$_{DIRBE}$ = (0.937 $\pm$ 0.003)J$_{2MASS}$ + 0.077 $\pm$ 0.012
($\chi$$^2$ = 4304
with $N$ = 4691)
and K$_{DIRBE}$ = (0.982 $\pm$ 0.004)K$_{2MASS}$ + 0.109 $\pm$ 0.004
($\chi$$^2$ = 3542
with $N$ = 5169).
No sources were flagged for discrepant photometry from 2MASS
that were not already flagged for a companion in the beam
or $\sigma$/$<$$err$$>$ $\ge$ 3 in a 2-week period.

DIRBE also provides information about near-infrared variability, which is
unavailable from 2MASS.  
In Table 7, we provide statistics on the DIRBE light curves and flagging
for the 791 sources in the 2MASS K $<$ 1.0 sample.  Of the
791 sources, 570 survive the filtering process at 2.2 $\mu$m (i.e., do
not have all their measurements removed by filtering), and are detected 
in a single DIRBE scan at S/N $\ge$ 
3.
Of these 570 sources, 271 are unflagged at 2.2 $\mu$m, and 
have more than 100 points left
in their light curves after filtering.  Of these, 90 (33$\%$) are
observably variable at 2.2 $\mu$m ($\ge$3$\sigma$ DIRBE
amplitudes).   

In Figure 9, 
we plot histograms of the observed DIRBE amplitudes at the six shortest
DIRBE wavelengths
for the sources in the K $<$ 1.0 subset.
This plot only includes unflagged sources with high S/N light curves
($>$3$\sigma$ per scan) with at
least 100 measurements remaining after filtering.  
The sizes of the bins at each wavelength are equal to the median
uncertainty in the amplitude at that wavelength.
These histograms, along with Table 7, show that 
the majority of the sources in this subset are not observably
variable in the DIRBE data.
For the sources that are variable,
the observed amplitudes are all less than 2.5 magnitudes
at 1.25 $\mu$m, less than 2 magnitudes at 2.2 $\mu$m and 3.5 $\mu$m,
and less than 1.5 magnitudes at 4.9, 12, and 25 $\mu$m.

\section{IRAS vs. DIRBE Variability}

In Figure 10a, we plot the DIRBE 12 $\mu$m flux density against
that from IRAS, for the unflagged $|$b$|$ $\ge$ 5$^{\circ}$ sources.
The DIRBE 25 $\mu$m flux densities for unflagged $|$b$|$ $\ge$ 
5$^{\circ}$ sources are plotted against the IRAS 25 $\mu$m flux
density in Figure 10b.
The errorbars plotted on the DIRBE flux densities 
are the standard deviations in the filtered light curves,
thus they include both measurement errors and intrinsic
variations.  The best-fit lines to these data
are $[$log~F$_{12}$$]$$_{DIRBE}$ = $[$0.979 $\pm$ 
0.014$]$$[$log~F$_{12}$$]$$_{IRAS}$
+ 0.010 $\pm$ 0.035 ($\chi$$^2$ = 431, N = 149), and
$[$log~F$_{25}$$]$$_{DIRBE}$ = $[$1.107 $\pm$ 
0.012$]$$[$log~F$_{25}$$]$$_{IRAS}$
- 0.135 $\pm$ 0.034 ($\chi$$^2$ = 220, N = 115).

Unlike the 2MASS Catalog, the IRAS Point Source Catalog provides
variability information, in the form of the IRAS VAR parameter
(see \citet{b88}).
Sources that were detected in at least two detectors 
during a single IRAS scan were considered
`seconds-confirmed'.  These seconds-confirmed 
detections were then `hours-confirmed'
by matching with 
detections in scans 
between 100 minutes to 36 hours later.
These hours-confirmed measurements were then compared
with other hours-confirmed detections
on longer timescales (weeks to months) (`weekly-confirmed').
Typically two or three hours-confirmed measurements
were obtained per object on the sky.
The IRAS variability parameter VAR was obtained
by comparing hours-confirmed measurements for 
sources with high or moderate quality fluxes at
both 12 and 25 $\mu$m.  For a source to be 
considered variable, fluxes at both 12 and 25 $\mu$m
must have increased or decreased significantly 
from one hours-confirmed sighting
to another (i.e., their variations must be correlated).
For the VAR parameter to be set to 99 percent probability, 
a 2.6$\sigma$ correlated change must have occurred at both
wavelengths; for VAR = 50 $-$ 98 percent probability,
a 0.9 $-$ 2.6$\sigma$ correlated change must have occurred.
This means that sources with very long periods
of variation may not have been flagged as variable in IRAS.
In addition, if two hours-confirmed observations were 
made during approximately same variability phase, the object may
not have been flagged as variable.  Furthermore, since the amount of
variation needed for a VAR $\sim$ 50$\%$ setting is
relatively low, VAR values in this range are
expected to be somewhat unreliable.

Because of the good temporal coverage of COBE,
the DIRBE variability parameters, particularly the amplitude
of variation compared to the uncertainty on the amplitude,
are expected to be more reliable indicators of variability
for high S/N unconfused objects.  DIRBE has the extra
advantage of additional shorter wavelength bands than IRAS.

In Figures 11a $-$ f, 
we compare 
the IRAS VAR parameter with the DIRBE amplitude of variation
at the six shortest DIRBE wavelengths
for the sources
in the \citet{smith03} sample (IRAS F$_{12}$ $\ge$ 150 Jy;
$|b$$|$ $\ge$ 5$^{\circ}$).
We have excluded sources with less than 5$\sigma$ DIRBE flux
densities at minimum light at the given wavelength, sources
that were flagged as possibly confused in the DIRBE database,
and sources which were flagged
in the IRAS Point Source Catalog as possibly being 
confused or in a region dominated by cirrus. 
We have
only included sources with high quality IRAS fluxes at
both 12 and 25 $\mu$m.

These plots show considerable scatter.
In particular, a number of stars that have low IRAS VAR
parameters have large amplitudes of variation in the DIRBE
database ($>$0.5 magnitude).   
As expected,
the IRAS satellite
was unsuccessful in detecting variability for some strongly 
variable stars.
DIRBE is also better at finding smaller variations than
IRAS.  This figure shows a number of sources that are
variable in DIRBE 
($\Delta$(mag)/$\sigma$($\Delta$(mag)) $>$ 3) with
low amplitude ($<$0.4 magnitudes), but have low IRAS VAR.
The DIRBE light curves of the stars variable in DIRBE but with 
low IRAS VAR have
been inspected by eye, and confirmed that they are indeed
variable in the DIRBE database.   Most of these stars
are well-known variable stars, including Mira, which
has an IRAS VAR parameter of only 3.   

Note, however, that 
the majority of VAR = 99 sources are strongly variable
in the DIRBE database.  In fact, most sources with VAR $\ge$ 80
are also variable in DIRBE.  Thus using the IRAS VAR parameter
to select highly variable sources (as in,
for example, \citet{akw93}) appears to be reasonably
reliable (albeit incomplete).

\section{Associations with Other Catalogs and Spectral Types}

To obtain associations with known optical sources and sources at other wavelengths,
the
DIRBE 
Catalog was cross-correlated with the SIMBAD database \citep{w96}.
The results are given in Table 8\footnote{Table 8 is only available
in electronic form, at the Astrophysical Journal Supplement
web site at http://www.journals.uchicago.edu/ApJ}.
The closest source listed in SIMBAD within 1$'$ was assumed to be the same source,
and the spectral type and object type of this source are given in Table 8.
SIMBAD sometimes has more than one listing for the same source,
if the source appeared in two different catalogs with slightly different positions
and different names.
Therefore, if a second SIMBAD source is listed 
within 5$''$ of the first source, that second
source and its SIMBAD object type and spectral type are also given in Table 8.
We also cross-correlated our list with the General Catalog of Variable Stars (GCVS;
\citep{GCVS}),
the New Catalog of Suspected Variables (NSV; \citep{NSV}),
and the NSV Supplement (NSVS; \citet{kd98}),
using search radii of 
60$''$, 30$''$, and 5$''$, respectively.
If a match occurred, the variability type and period are included
in Table 8, if available.
Table 8 also includes
IRAS spectral types 
from \citet{kvb97}
for the IRAS sources.
In addition, 
we compared our list with the 
\citet{c01}
OH/IR star compilation, which is listed by IRAS name.
DIRBE Catalog sources that are in that list are noted in Table 8, and the
OH expansion velocity is included.

Although the SIMBAD classifications are inhomogeneous and incomplete,
they provide some rudimentary statistical information about the types
of sources contained in the DIRBE Catalog.
In Table 9, we provide statistics on types as a function of wavelength, for the sources
with S/N per scan $\ge$ 3 at that wavelength.  
The sources have been divided into 18 groups, based
on their SIMBAD spectral type and object type, as well as their
variability type from the GCVS, NSV, or NSVS, if available.
Stars with optical types M, K, S, and C are divided into two groups: those with
variability types associated with the AGB (Mira, SRa, SRb, Lb),
and those not previously classified into one of those variability types.
Objects identified in SIMBAD, the GCVS, the NSV, or
the NSVS as young stellar objects, Orion variables, Herbig Ae/Be
stars, T Tauri stars, H~II regions, or Herbig-Haro objects are lumped
together as star formation sources.
Sources listed in SIMBAD as post-AGB objects are also separated out,
as are planetary nebulae, possible planetary
nebulae, and galaxies.  Stars with optical types of O, B, A, F, and G
not included in one of these sets are tabulated separately.
A similar listing for the 2MASS K $<$ 1.0 subsample is presented in
Table 10.
This near-infrared-bright subset has no known star formation
sources, post-AGB objects, or planetary nebulae, but 
contains mainly late-type stars.

In addition to providing the number of sources
of each type detected at each wavelength at S/N $\ge$ 3 per scan,
in Tables 9 and 10 
we give statistics on the number of these that are unflagged and have
at least 100 measurements left after filtering.   We also include the
number of unflagged sources with $\ge$ 100 measurements
that are variable with amplitude/$\sigma$-amplitude
$\ge$ 3.
Table 9 shows that, of the 2352 unflagged high S/N 4.9 $\mu$m sources,
984 were previously classified 
as Mira, SRa, SRb, or Lb.  Of these, 152 (15$\%$) are observably
variable in the DIRBE database. 
Of the remaining 1368 not previously classified as one of these variability
types, 70 (5$\%$) are variable in DIRBE at the 3$\sigma$ level.
At 12 $\mu$m, 85 of 311 (27$\%$) unflagged high S/N sources
are variable; 42 of these (49$\%$) 
are not previously classified as Mira, SRa, SRb, or Lb.
Note that, of the unflagged carbon stars not previously
classified as one of these variability types, more than half 
of the stars that are detected at 4.9, 12, or
25 $\mu$m are variable at these wavelengths.

\section{The Brightest Catalog Sources }

In Tables 11 $-$ 16, the brightest 10 sources in the DIRBE Catalog at 
each of
the six shortest wavelengths are given.
Although these are the brightest sources in the Catalog, they are 
not necessarily the brightest sources in the sky because filtering
eliminates some sources and because of incompleteness in the Galactic plane.
The characteristics of the brightest sources change with wavelength.
At the shortest wavelengths, optically-bright semi-regulars and non-variable
K and
M stars dominate.  
At the longer wavelengths, more evolved dusty stars
and star formation regions become more important.

Some of these extremely bright sources are clearly variable in DIRBE.
For example, see
the 2.2 $-$ 60 $\mu$m DIRBE light curves for W Hya
shown in Figure 6 
and
the DIRBE light curves of 
CW Leo, 
$\alpha$ Ori, 
VY CMa, and L$_2$ Pup presented in 
Figure 1 and
\citet{smith03}.

At 12 and 25 $\mu$m there is considerable confusion
in the vicinity of star formation regions, such as that
in Orion.
This demonstrates the need for caution in using flagged photometry
from the Catalog.

\section{DIRBE Variability of AGB Stars}

As shown in Table 9, the vast majority of objects in the DIRBE Catalog
with known optical spectral types are late-type stars (spectral types
M, K, S, and C).
Of these, more than half are known variables with variability
types associated with the AGB.
In \citet{smith02} and \citet{smith03}, we showed
that 
the amplitudes of variation 
for these types of stars
tend to decrease with increasing wavelength.  Further, the amplitudes
increased along the sequence SRb/Lb $-$$>$ SRa $-$$>$ Mira.
The DIRBE Catalog contains a significantly larger number of
sources than these earlier studies, so we have repeated these analyses 
with the full Catalog sample.

In Table 17, we provide the mean DIRBE amplitudes of variation at 
each wavelength
{\it 
$<$$\Delta$(mag)$>$ }
for the known Miras, SRa, SRb, and Lb stars in the DIRBE Catalog.  We have
separated the stars into groups according to whether they are 
oxygen-rich (M stars and/or IRAS LRS types E or A; see \citet{kvb97}),
carbon-rich (optical types C, N or R, or IRAS type C), 
optical type S (C/O ratio $\sim$ 1),
or known OH/IR stars (often also
optical type M and/or IRAS type A or E).
We also include the average uncertainty in the amplitude,
$<{\sigma}({\Delta}(mag))>$.
In constructing Table 17, we only included unflagged stars with $\ge$5$\sigma$
DIRBE flux densities at minimum light.

In Table 18, 
we provide the mean
ratios of the amplitudes at adjacent wavelengths
for the different classes of objects.
The quoted uncertainties in this Table are the standard deviations
of the ratios for each class.
Table 18 excludes flagged stars and 
stars with flux densities less than 5$\sigma$ at minimum and $\Delta$(mag)/$\sigma$($\Delta$(mag)) $<$
5
at either wavelength.
These results are shown graphically in Figure 12, where we plot
the mean amplitude ratio for two adjacent wavelength bands as a function
of the shorter wavelength.
Table 18 and Figure 12 show that,
on average,
there is a decrease in amplitude
with wavelength, with the exception of the 4.9 $\mu$m/12 $\mu$m amplitudes
for the oxygen-rich Miras, the 2.2 $\mu$m/3.5 $\mu$m amplitudes for the carbon Miras
and the oxygen-rich SRb stars, and the 1.25 $\mu$m/2.2 $\mu$m
amplitudes for the oxygen-rich SRa stars.
There is, however, a lot of scatter in the amplitude ratios from star
to star;  although the average ratio tends to decrease with wavelength,
for some individual stars the ratio increases or remains constant.
This general trend is 
consistent with earlier DIRBE results with a smaller sample
\citep{smith03} as well as ground-based studies
\citep{harvey74}.

Such amplitude decrements with wavelength are predicted by theoretical
models of AGB stars \citep{lebetre88, winters94, winters00}.
A decrease in amplitude with wavelength
means that the dust shell is redder
at stellar minimum than at maximum.
At any given radius in a circumstellar shell, the average dust temperature
is expected to be highest at stellar maximum, and the radii that dust condensation
and evaporation
occur are largest at stellar maximum.
Assuming a roughly 1/r$^2$ density distribution for the dust in the shell and
integrating over all of the dust in the shell, the optical depth of the shell is therefore
highest at
minimum, while the average effective dust temperature
for the shell is lowest. This produces a redder infrared spectrum at minimum.

\section{Massive Evolved Stars}

In \citet{smith03}, we presented the DIRBE light curves
of the supergiants $\alpha$ Ori and VY CMa.
The dusty object VY CMa, the fourth brightest object in
the DIRBE Catalog at 25 $\mu$m (Table 16), 
is variable in the mid-infrared, but
not as variable as a typical Mira. 
Another object that may be similar to VY CMa is NML Cyg (V1489 Cyg)
\citep{nml65}, 
also an OH/IR star and
the thirteenth brightest 25 $\mu$m DIRBE Catalog source.
NML Cyg may also be a supergiant \citep{johnson67, mj83}.
NML Cyg is clearly variable in the DIRBE database (see Figure 13a),
with observed amplitudes of 
0.30 $\pm$ 0.03, 0.23 $\pm$ 0.06, and 0.20 $\pm$ 0.06 
magnitudes at 3.5, 12, and 25 $\mu$m, respectively.
These amplitudes are somewhat larger than those of VY CMa,
but smaller than typical values for Miras (see Table 17).

\citet{monnier97} published a 10.2 $\mu$m light curve for 1980 $-$ 1995 
for NML Cyg, and found a period of $\sim$940 days.
Their photometry is consistent within the uncertainties
with that of DIRBE for the time period
in common.  The DIRBE photometry and amplitudes are also reasonably
consistent with the older data of \citet{harvey74} and \citet{strecker75},
as well as 
the IRAS time-averaged 
flux densities (from the xscanpi software)
of
F$_{12}$ = 5461 Jy and F$_{25}$ = 4065 Jy, and the MSX photometry
\citep{e99}.

\section{Planetary Nebulae and Post-AGB Objects}

Only 8 sources detected by DIRBE
at the $\ge$3$\sigma$ level per measurement
at 4.9 $\mu$m are classified in SIMBAD
as post-AGB objects, while 4 are listed as planetary
nebulae or suspected planetary nebulae.
None of these sources
were found to be
unconfused and variable at the $\ge$3$\sigma$ level in DIRBE at
any wavelength (Table 9). In Figure 13b, we show a few light curves
for the post-AGB object RAFGL 2688 (the Egg Nebula), which is not
variable in DIRBE.

\section{An Object in Transition: OH 231.8+4.2 (QX Pup)}

One of the most intriguing objects in the DIRBE Catalog
is OH 231.8+4.2 (QX Pup), which apparently is in the process
of changing from an AGB star to a post-AGB object \citep{meakin03}.
OH 231.8+4.2 is a bipolar reflection nebula \citep{reipurth87}
with an embedded M9III star \citep{cohen81}.
This star is known to be variable 
with an amplitude of approximately 2 magnitudes at 2.2 $\mu$m
and a pulsation period of 700 days \citep{kast92},
typical of Miras.  This is thus a very unusual source, in a 
brief and rarely observed stage of development: a pulsating AGB star 
inside of a proto-planetary nebula.

OH 231.8+4.2 is variable in the DIRBE database, with an amplitude of
0.29 $\pm$ 0.07 magnitudes at 25 $\mu$m 
(see Figure 13c).   
This confirms the high IRAS VAR of 99 for this source.
The DIRBE 25 $\mu$m amplitude 
for OH 231.8+4.2
is less than the average for the known 
Miras in the DIRBE Catalog (Table 17), which typically have amplitudes 
of 
0.5 magnitudes at 25 $\mu$m, however, DIRBE may not have observed a 
full pulsation cycle for this star.  OH 231.8+4.2 is also 
variable at longer wavelengths,
with a measured amplitude of 
approximately 0.32 magnitudes 
at 850 $\mu$m \citep{j02}.

OH 231.8+4.2 is one of the reddest evolved stars in the DIRBE Catalog.
In Figure 14, we plot the IRAS flux ratios
for the 43 unconfused sources in the DIRBE Catalog
with IRAS colors
F$_{25}$/F$_{12}$ $\ge$ 1.2 and
F$_{60}$/F$_{25}$ $\le$ 2.5, and high quality IRAS fluxes
at 12, 25, and 60 $\mu$m.
These colors are in the range expected for very late
AGB stars and post-AGB objects \citep{vdvh88}.
Using types from SIMBAD and additional information from the literature,
in Figure 14
we use different symbols to identify young stars,
post-AGB objects,
planetary nebula,
and 
OH/IR stars not identified as post-AGB objects.
For comparison,
we also mark the location of 
the post-AGB 
Egg Nebula (RAFGL 2688), based on its DIRBE data, and the
variable supergiant NML Cyg, based on IRAS xscanpi results.
Of these 45 objects, eight were found to be variable in
the DIRBE database.  These are distinguished by filled symbols 
in Figure 14.

In Figure 14, we also plot
the regions 
defined by \citet{vdvh88}.
As discussed by \citet{vdvh88},
sources in 
Regions II, IIIa, and IIIb are mainly AGB stars,
with increasing circumstellar shell optical depth along this sequence.
Region IV contains very late-stage AGB stars as well as some planetary
nebulae.
We also plot the theoretical 
evolutionary track for AGB stars of \citet{b87}
in Figure 14.
Stars get redder as they evolve, moving to the 
upper right on the IRAS color-color plot.
After the stars evolve into 
region IV of the IRAS color-color plot, 
they move into
Region V, which contains mainly non-variable objects,
including planetary nebulae and post-AGB objects
\citep{vdvh88}.
OH 231.8+4.2 is
the reddest source in this plot,
lying
in the extreme upper right of the color-color plot, to the right of 
region V, far from the standard AGB evolutionary track, and far
from the other sources found to be variable in DIRBE.

\section{Star Formation Regions and
Other O, B, A, F, and G Stars}

As shown in Table 9, 47 of the sources detected
at the 3$\sigma$ level at 4.9 $\mu$m in a single DIRBE scan
are associated with star formation regions.
Confusion is a major problem for the star formation sources
in the Catalog; only 9 of these sources are unflagged.
None
of these are variable in DIRBE
at a $\ge$3$\sigma$ level.

Table 9 shows that 150 of the high S/N unflagged 4.9 $\mu$m sources in the
DIRBE Catalog are O, B, A, F, or G stars, but are not previously
identified as being associated with star formation regions or post-AGB
objects.
Of these sources, none are variable in DIRBE.

\section{Galaxies}

As noted above, the DIRBE Catalog selection criteria are biased against
galaxies, since we are only selecting sources at wavelengths $\le$ 25 $\mu$m
and are biased against extended objects.
Only two galaxies are included in the final DIRBE Catalog:
M82 and NGC 253.  Their IRAS flux densities fit our selection
criteria, and they are detected 
at $\ge$3$\sigma$
in a single DIRBE scan at 12 and 25 $\mu$m.  As expected, their light
curves show no evidence for variations.
NGC 253 is also detected at 1.25 and 2.2 $\mu$m with average flux densities of
20 $\pm$ 4 Jy and 19 $\pm$ 6 Jy per measurement, consistent with values in
the 2MASS Large Galaxy Atlas\footnote{Available 
at http://irsa.ipac.caltech.edu/applications/Gator/.}
\citep{jarrett03}.
By co-addition over the entire DIRBE mission, it is possible to detect
additional galaxies at wavelengths of $\ge$60 $\mu$m
(see \citet{ons98}).
Such co-addition is beyond the scope of the current Catalog.

\section{Summary}

From the archival COBE DIRBE database,
we have constructed a DIRBE Point Source Catalog containing
11,788 sources detected at the 3$\sigma$ level per measurement
at at least one DIRBE wavelength.  This catalog
was created using an input list of 21,335 IRAS/MSX and 2MASS sources.
We have flagged sources likely to be confused, based on other
infrared catalogs and inspection of the DIRBE light curves, and 
have collected information about the DIRBE sources from other catalogs.
We compare the DIRBE photometry with that of 2MASS
for the near-infrared-bright sources in the DIRBE Catalog, 
and show that
the DIRBE photometry
is more precise.  We compare the DIRBE variability parameters
with the IRAS VAR parameter, and show that DIRBE detected
variability in a number of sources with low IRAS VAR.
We also discuss a few unusual objects in the Catalog, including
the peculiar bipolar nebula OH 231.8+4.2, likely evolving from
the AGB into a post-AGB object.

\vskip 0.1in

We would like to thank the COBE team for making this project
possible.  We are especially grateful to the DIRBE 
Principal Investigator Michael Hauser.
The DIRBE Calibrated Individual Observations data product was
developed by the COBE Science Working Group and provided by the National
Space Science Data Center at NASA's Goddard Space Flight Center.
We appreciate the work of Nils Odegard, who wrote the first version of
the DIRBE light curve extraction routine.
We also thank the referee, David Leisawitz, for helpful comments
that significantly improved this paper.
We also thank Anatoly Miroshnichenko and Mark
Giroux for helpful communications, and J. M. Houchins for computer
support.
This research has made use of the SIMBAD database,
operated at the CDS, Strasbourg, France, as well
as NASA Astrophysics Data System
at the Harvard-Smithsonian Center for Astrophysics,
the VizieR service at the CDS, Strasbourg, France,
and the Infrared Science Archive, operated by
the Jet Propulsion Laboratory, California Institute
of Technology.
We have also made use of the electronic
versions of the General Catalogue
of Variable Stars, the New Suspected Variable Stars
Catalogue, and the Supplement to this catalogue, provided
by the Sternberg Astronomical Institute at Moscow
State University.
This research was funded by
National Science Foundation POWRE
grant AST-0073853 and NASA LTSA grant NAG5-13079.

\vfill
\eject

%% Generally speaking, only the figure captions, and not the figures
%% themselves, are included in electronic manuscript submissions.
%% Use \figcaption to format your figure captions. They should begin on a
%% new page.

\clearpage

%% No more than seven \figcaption commands are allowed per page,
%% so if you have more than seven captions, insert a \clearpage

%% There must be a \figcaption command for each legend. Key the text of the
%% legend and the optional \label in curly braces. If you wish, you may
%% include the name of the corresponding figure file in square brackets.
%% The label is for identification purposes only. It will not insert the
%% figures themselves into the document.
%% If you want to include your art in the paper, use \plotone.
%% Refer to the on-line documentation for details.

\begin{figure}
\epsscale{0.8}
\plotone{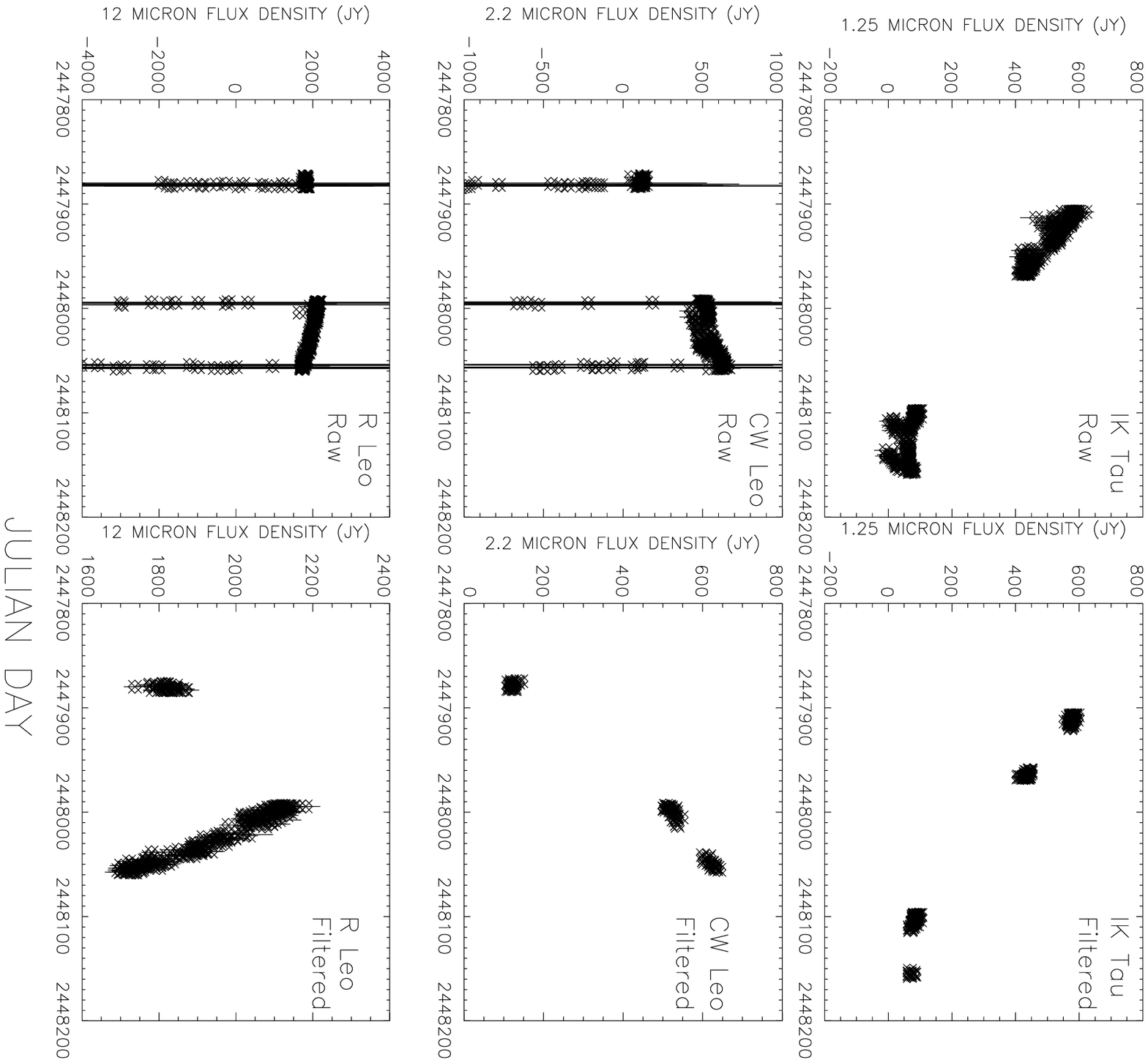}
\caption{\small 
Some examples of light curve filtering.
a) The raw 1.25 $\mu$m DIRBE light curve of IK Tau.
b) The filtered 1.25 $\mu$m DIRBE light curve of IK Tau.
c) The raw 2.2 $\mu$m light curve for CW Leo.
d) The filtered 2.2 $\mu$m light curve for CW Leo.  
e) The raw 12 $\mu$m light curve of R Leo.
f) The filtered 12 $\mu$m light curve of R Leo.
}
\end{figure}

\begin{figure}
\epsscale{0.8}
\plotone{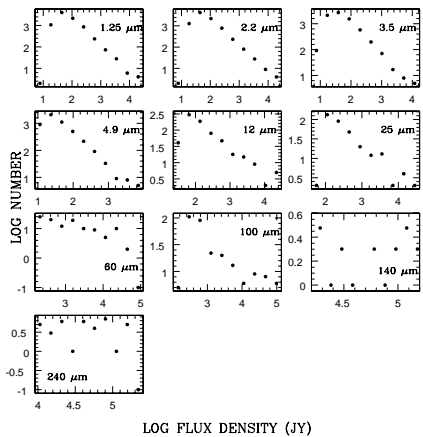}
\caption{\small 
Log N $-$ log S plots made from the DIRBE Point Source Catalog, for high
Galactic latitude sources ($|$b$|$ $\ge$ 5$^{\circ}$).
}
\end{figure}

\begin{figure}
\epsscale{0.8}
\plotone{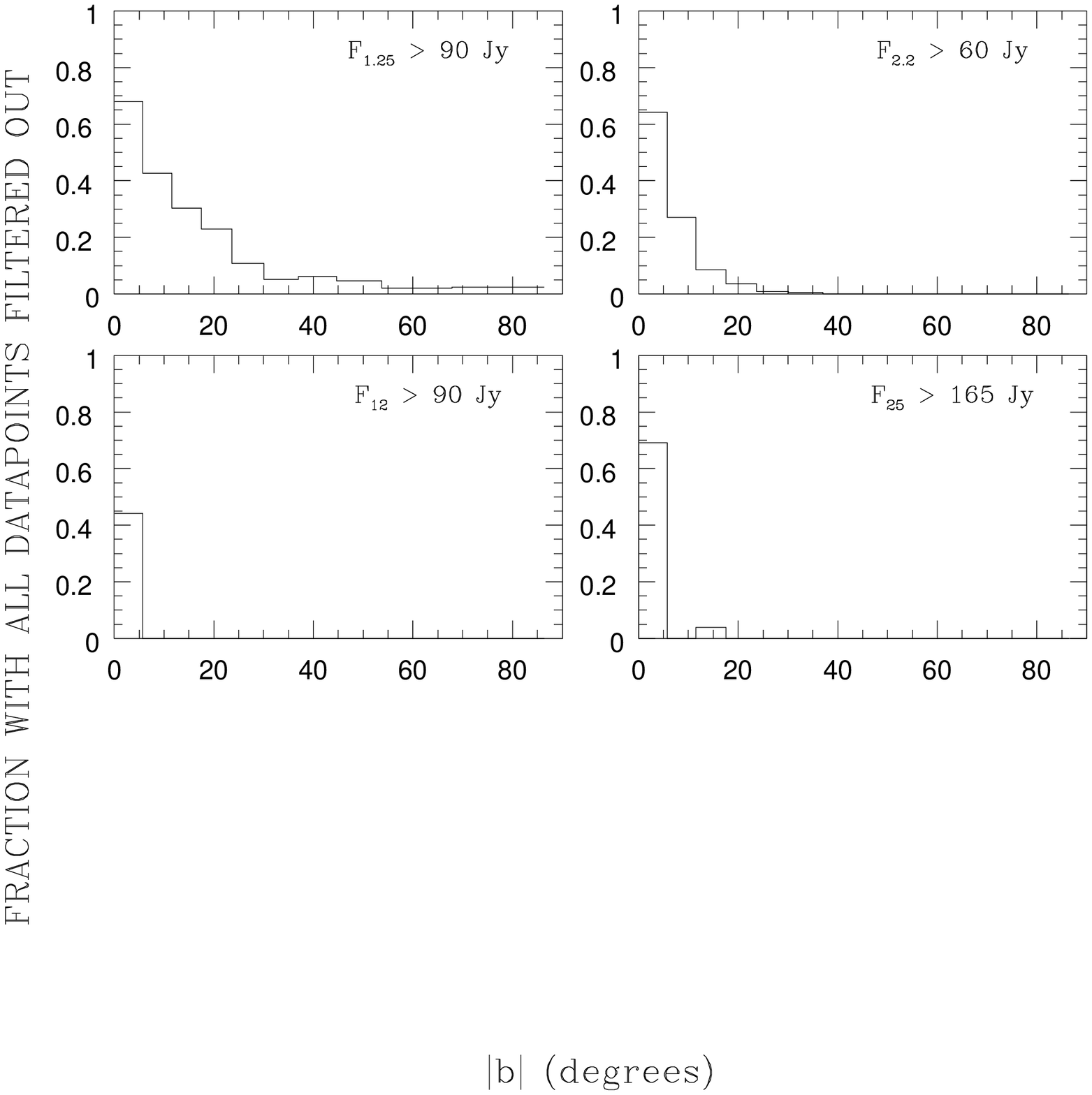}
\caption{\small 
The distribution with Galactic latitude 
of
the fraction of 1.25, 2.2, 12, and 25 $\mu$m sources 
with flux densities 
from 2MASS, IRAS, or MSX 
above the Catalog input list nominal completeness limits
that have all their DIRBE measurements removed by filtering.
The bins cover equal sky area.
}
\end{figure}

\begin{figure}
\epsscale{0.8}
\plotone{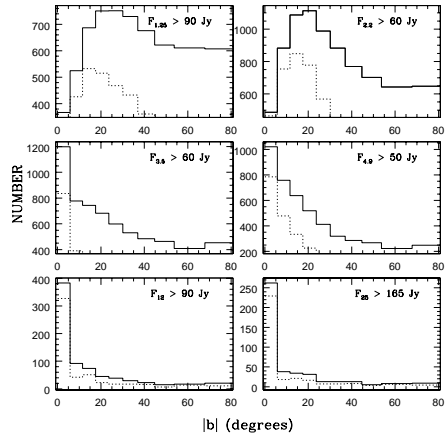}
\caption{\small 
Solid line: A histogram of the number of sources in the DIRBE Catalog
with DIRBE photometry above the Catalog input list nominal
completeness limits.  
Dotted line: a similar histogram for the flagged sources.
The bins cover equal sky areas.
}
\end{figure}

\begin{figure}
\epsscale{0.8}
\plotone{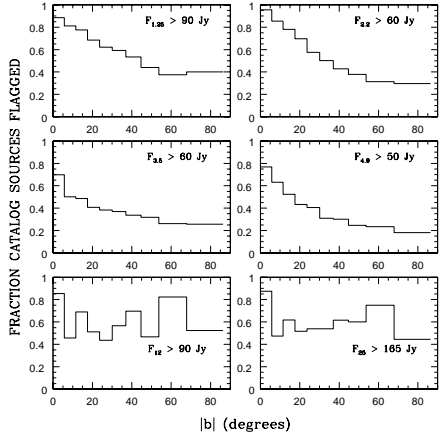}
\caption{\small 
The fraction of flagged sources above the nominal completeness limits
as a function of Galactic latitude.
The bins cover equal sky area.
}
\end{figure}

\begin{figure}
\epsscale{0.8}
\plotone{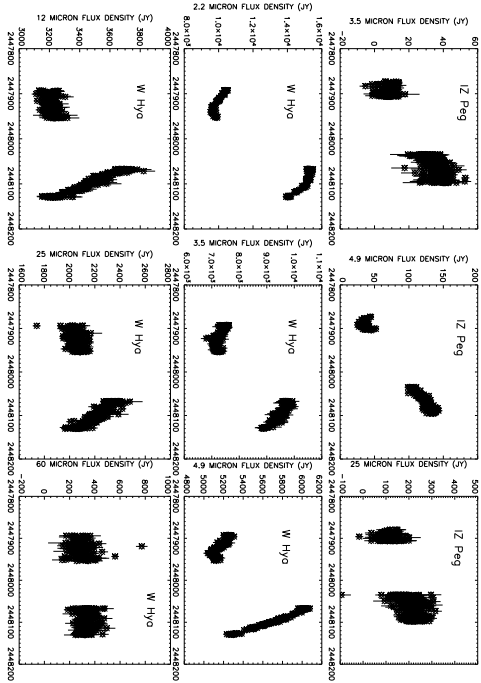}
\caption{\small 
Some example light curves.
Top row:
DIRBE light curves at 3.5, 4.9, and 25 $\mu$m for IZ Peg.
Middle and bottom row:
DIRBE light curves at 2.2, 3.5, 4.9, 12, 25, and 60 $\mu$m for W Hya.
}
\end{figure}

\begin{figure}
\epsscale{0.8}
\plotone{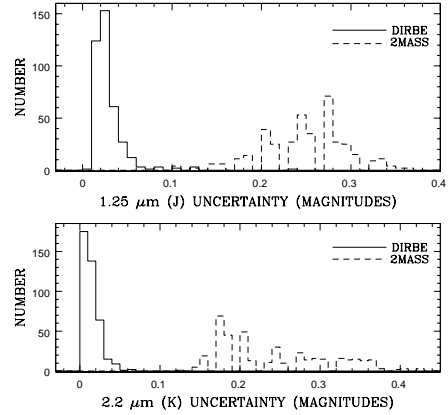}
\caption{\small 
The distribution of uncertainties in the J (1.25 $\mu$m) and K
(2.2 $\mu$m) magnitudes in DIRBE and 2MASS for the 791 brightest 2.2 $\mu$m 
sources in the sky (2MASS K $\le$ 1.0 
(F$_{2.2}$ $\ge$ 266 Jy)).
}
\end{figure}

\begin{figure}
\epsscale{0.8}
\plotone{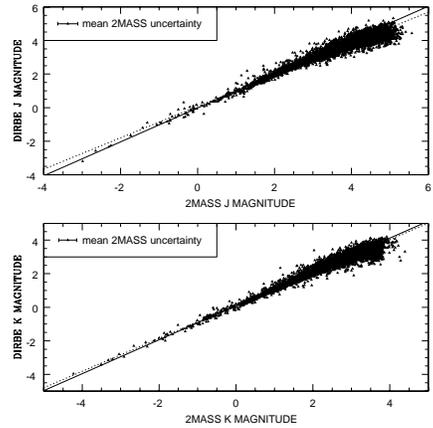}
\caption{\small 
Top panel: A comparison of the 2MASS J magnitudes with those from DIRBE for the
unflagged $|$b$|$ $\ge$ 5$^{\circ}$ sources.
Bottom panel: The 2MASS K magnitudes plotted against those from DIRBE, for unflagged
sources.
For clarity, errorbars are not included in the plot.  Instead, the mean
2MASS uncertainty is shown in the upper left.   The mean DIRBE uncertainty
is smaller than the size of the datapoints.
The best-fit lines as given in the text are shown; the solid lines are the 
best-fit for the data with 2MASS K $<$ 1, while the dotted lines are for the full Catalog.
Sources brighter than J $\sim$ 4.5 and K $\sim$ 3.5 are saturated in 2MASS
\citep{c03}.
}
\end{figure}

\begin{figure}
\epsscale{0.8}
\plotone{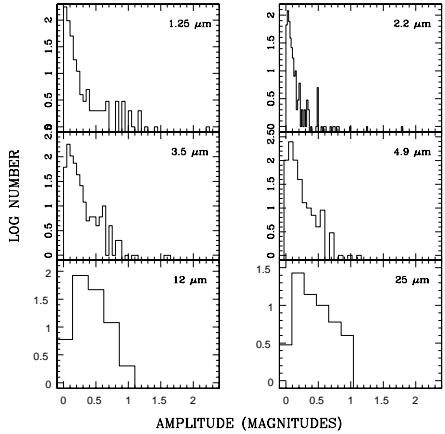}
\caption{\small 
Histograms of the 1.25 $\mu$m $-$ 25 $\mu$m DIRBE amplitudes
for the 791 sources with 2MASS K $\le$ 1.0
(F$_{2.2}$ $\ge$ 266 Jy).
This plot includes only unflagged sources, with filtered
light curves containing $\ge$100 measurements.  
The bin sizes are set equal to the median uncertainties in
the amplitude at each wavelength, which are 0.05, 0.02, 0.05,
0.07, 0.24, and 0.19 magnitudes at
1.25, 2.2, 3.5, 4.9, 12, and 25 $\mu$m,
respectively.
}
\end{figure}

\begin{figure}
\epsscale{0.8}
\plotone{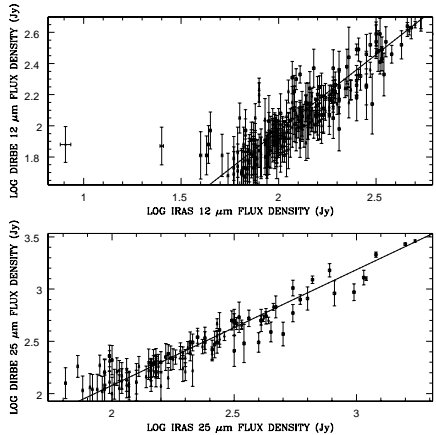}
\caption{\small 
A comparison of the DIRBE 12 and 25 $\mu$m flux densities with
those at IRAS, for unflagged $|$b$|$ $\ge$ 5$^{\circ}$ sources.
The best-fit lines as discussed in the text are shown. 
The very discrepant point to the left in the 12 $\mu$m plot
is a nebula in the Large Magellanic Cloud, thus it may
be extended or confused.
The open squares are points with $\Delta$(mag)/$\sigma$($\Delta$(mag))
$\ge$ 3 in the DIRBE Catalog.
}
\end{figure}

\begin{figure}
\epsscale{0.8}
\plotone{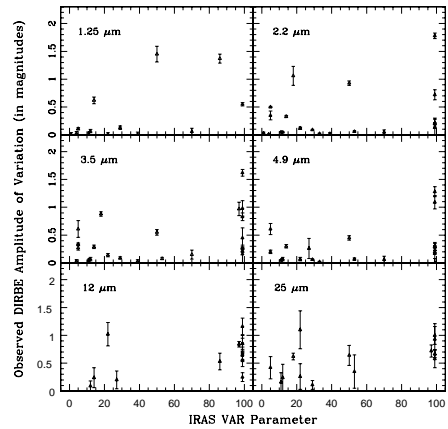}
\caption{\small 
A comparison of the IRAS VAR parameter with the DIRBE amplitudes
of variation at the six shortest DIRBE wavelengths, for the
207 brightest
high Galactic latitude ($|b$$|$ $\ge$ 5$^{\circ}$)
12 $\mu$m sources (F$_{12}$ $\ge$ 150 Jy).
Sources that have been flagged in DIRBE or are low S/N ($\le$5 at minimum)
at a wavelength have been excluded
at that wavelength.
We also excluded sources flagged in IRAS as being possibly confused 
or in a region dominated by cirrus (IRAS CONFUSE flag $\ne$ 0, CIRR1 $>$ 3,
or CIRR2 $>$ 4).  We have only included sources with high quality
IRAS fluxes at both 12 and 25 $\mu$m (FQUAL12 and FQUAL25 = 3).
}
\end{figure}

\begin{figure}
\epsscale{0.8}
\plotone{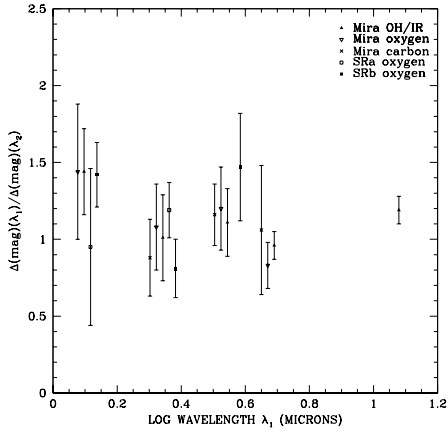}
\caption{\small 
The ratio of the observed DIRBE amplitude at a given wavelength $\lambda$$_1$ to
the amplitude at the next longest DIRBE wavelength $\lambda$$_2$, plotted as  
a function of $\lambda$$_1$.
This plot excludes flagged stars and stars detected with less than S/N = 5 at
minimum in the weekly-averaged light curves at both wavelengths.
It also excludes stars with $\Delta$(mag)/$\sigma$($\Delta$(mag)) $<$ 5 at both
wavelengths.  Only classes of stars with at least 3 stars that fit these criteria 
at a given wavelength are plotted.
For clarity, some datapoints have been shifted slightly in wavelength.
}
\end{figure}

\begin{figure}
\epsscale{0.8}
\plotone{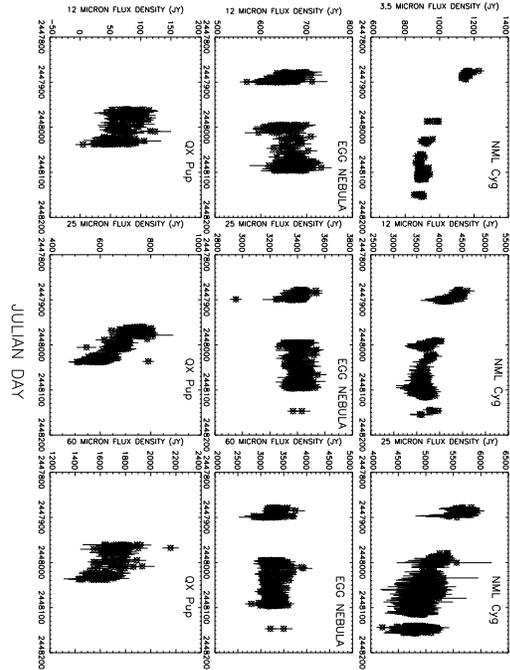}
\caption{\small 
Some more example light curves.
a) - c)
DIRBE light curves at 3.5, 12, and 25 $\mu$m for NML Cyg.
d) - f)
DIRBE light curves at 12, 25, and 60 $\mu$m for the Egg Nebula (RAFGL 2688).
g) - i)
DIRBE light curves at 12, 25, and 60 $\mu$m for QX Pup (OH231.8+4.2).
}
\end{figure}

\begin{figure}
\epsscale{0.8}
\plotone{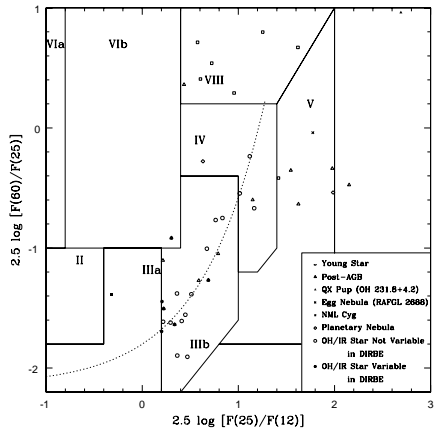}
\caption{\small 
The IRAS color-color plot for the 43 unconfused objects in the
DIRBE Catalog with a high S/N ($\ge$3) DIRBE light curve with
at least 100 unfiltered data points at at
least one wavelength, with IRAS
F$_{25}$/F$_{12}$ $\ge$ 1.2 and F$_{60}$/F$_{100}$ $\ge$ 2.5,
and with 
high quality IRAS Point Source Catalog fluxes at 12, 25, and 60 $\mu$m. 
Sources at all Galactic latitudes are included.
Filled symbols represent objects which
are variable in the DIRBE database; open symbols are objects
that are not variable.  The filled triangle in the upper right 
is QX Pup (OH 231.8+4.2).
The cross marks the location based on the DIRBE data 
of the Egg Nebula (RAFGL 2688), a post-AGB
object not variable in DIRBE, while the filled square shows the
location of the supergiant NML Cyg, based on IRAS xscanpi results.
The regions labeled with Roman numerals are from van der Veen
$\&$ Habing (1988).  The dashed line is the Bedijn (1987) 
evolutionary track for AGB stars.
}
\end{figure}

\clearpage

%% manuscript produces a one-column, double-spaced document:

%% preprint2 produces a double-column, single-spaced document:

\begin{deluxetable}{|c|c|c|r|r|r|r|r|r|r|r|r|r|r|r|}
\tabletypesize{\scriptsize}
\rotate
\tablecaption{DIRBE PHOTOMETRY\label{tbl-1}}
\tablewidth{0pt}
\tablehead{
\colhead{NAME}&
\colhead{2MASS}&
\colhead{IRAS/MSX}&
\multicolumn{6}{c}{1.25 $\mu$m}&
\multicolumn{6}{c}{2.2 $\mu$m}\\
\colhead{J2000 COORDINATES} & 
\colhead{NAME} & 
\colhead{NAME} & 
\colhead{F$_{\nu}$}   & \colhead{$\sigma$}   &
\colhead{$<$err$>$}&
\colhead{$\Delta$mag} &
\colhead{$\sigma$($\Delta$mag)} &
\colhead{N}&
\colhead{F$_{\nu}$}   & \colhead{$\sigma$}   &
\colhead{$<$err$>$}&
\colhead{$\Delta$mag} &
\colhead{$\sigma$($\Delta$mag)} &
\colhead{N}
\\
\colhead{}&
\colhead{}&
\colhead{}&
\colhead{(Jy)}&
\colhead{(Jy)}&
\colhead{(Jy)}&
\colhead{}&
\colhead{}&
\colhead{}&
\colhead{(Jy)}&
\colhead{(Jy)}&
\colhead{(Jy)}&
\colhead{}&
\colhead{}&
\\
}
\startdata
D00000657P2553112& 00000657+2553112 & IRAS23575+2536 &             162.1 &     47.4  &     12.1 &  0.82 &  0.13 &  180 &    233.9 &   
  38.3  &      9.1  & 0.49 &  0.06 &  286 \\
D00000690P2014145& 00000690+2014145  & &                            47.8 &      6.9  &     6.2 &  0.35 &  0.22 & 341 &     55.0 &   
   6.5 &      5.5  & 0.30 &  0.16 & 404 \\    
D00001353P5541579& 00001353+5541579   &  &                           84.5  &     26.6 &     15.0 &   0.73 &   0.35 & 289  &    77.8 &    
  48.3 &     24.0 & -99.90 & -99.90 & 350 \\
D00001815P6021016 & 00001815+6021016   &   &                          324.8  &    52.1 &     25.0 &  0.22 &  0.13 & 265   &   462.0  &  
  68.2 &     32.3 &  0.29 &  0.12 & 332 \\
D00010244P3830145 & 00010244+3830145 & IRAS23584+3813  &        83.2  &    11.2  &      7.0  &  0.36 &   0.13 & 378 &    108.7 &   
   9.0 &      6.3 &  0.21 &  0.09 & 385 \\
\enddata
\tablenotetext{a}{NOTE:-Table 1 is presented in its entirety, including additional columns, in the electronic
edition of the Astrophysical Journal Supplement.  A portion is shown here for guidance regarding
its form and content. }
\end{deluxetable}

\clearpage

\begin{deluxetable}{|c|c|}
\tabletypesize{\scriptsize}
\tablecaption{CONFUSION FLAGS IN THE DIRBE POINT SOURCE CATALOG\label{tbl-2}}
\tablehead{
{\bf FLAG} & {\bf DEFINITION }
\\
}
\startdata
1 & If set to 1, there is another source within 0.5$^{\circ}$ of the target source\\
 &above 25, 20, 20, 10, 30, 55, 320, and 765 Jy at 1.25, 2.2, 3.5, 4.9, 12, 25, 60, and 100 $\mu$m, respectively. \\
 &  This flag is not set at 140 and 240 $\mu$m.  \\
\hline
2 & If set to 1, the DIRBE photometry at minimum (in the weekly-averaged light curve) is greater than that of \\
 &2MASS or IRAS/MSX by more than 3$\sigma$, or the DIRBE photometry at maximum is less than that of 2MASS or \\
 &IRAS/MSX
 by 
more than 3$\sigma$. \\
 &  This flag is not set at 3.5, 4.9, 140, and 240 $\mu$m. \\
\hline
3 & If set to 1, the rms/$<$errorbar$>$ for a 2-week period is greater than 3, \\
 &suggesting confusion from an unknown source that is not completely corrected by filtering. \\
\hline
\enddata
\end{deluxetable}

\clearpage

\begin{deluxetable}{|c|c|c|c|c|c|c|c|c|c|c|}
%\tabletypesize{\scriptsize}
\tabletypesize{\tiny}
\rotate
\tablecaption{STATISTICS ON THE DIRBE POINT SOURCE CATALOG: FULL SAMPLE\label{tbl-3}}
\tablewidth{0pt}
\tablehead{
\colhead{}&
\colhead{1.25 $\mu$m}&   
\colhead{2.2 $\mu$m}&   
\colhead{3.5 $\mu$m}&   
\colhead{4.9 $\mu$m}&   
\colhead{12 $\mu$m}&   
\colhead{25 $\mu$m}&   
\colhead{60 $\mu$m}&   
\colhead{100 $\mu$m}&   
\colhead{140 $\mu$m}&   
\colhead{240 $\mu$m}
\\
}
\startdata

Average Uncertainty per&&&&&&&&&&\\
Datapoint in Raw Light Curves&33 Jy&38 Jy& 27 Jy&21 Jy &107 Jy & 256 Jy& 1567 Jy& 3207 Jy & 8100 Jy & 4510 Jy\\
\hline
Average Uncertainty per&&&&&&&&&&\\
Datapoint in Filtered Light Curves&
30 Jy&7 Jy&17 Jy&7 Jy&18 Jy&32 Jy&118 Jy &468 Jy& 8100 Jy & 4510 Jy\\
\hline
Estimated Completeness Limit for $|$b$|$ $\ge$ 5$^{\circ}$
&&&&&&&&&&\\
for DIRBE Catalog Input List&90 Jy& 60 Jy & 60 Jy&50 Jy&90 Jy& 165 Jy& $-$ & $-$ & $-$ & $-$ \\
\hline
Number Detected with&&&&&&&&&&\\
S/N $>$ 3 per Datapoint&6328&8203&6338&4688&894&502&448&395&136&209\\
\hline
Number Flagged for &&&&&&&&&&\\
Companion in Beam&5943&9825&1126&4848&3907&3262&2513&2674&$-$&$-$\\
\hline
Number Flagged for &&&&&&&&&&\\
Discrepancy with Other Photometry&198&152&$-$&$-$&308&173&249&186&$-$&$-$\\
\hline
Number Flagged for
&&&&&&&&&&\\
2-week rms/$<$error$>$ $\ge$ 3&
1171&1997&7274&1989&2242&2325&3889&3940&2774&3322\\
\hline
Number Unflagged
$>$3$\sigma$ Detections
&&&&&&&&&&\\
per Datapoint with N $\ge$ 100&
2521&3361&3513&2352&311&145&56&77&44&47\\
\hline
Number Unflagged $>$3$\sigma$ Detections per&&&&&&&&&&\\
Datapoint with $>$3$\sigma$ Amplitudes and N $\ge$ 100&
126&223&310&222&85&42&0&2&1&0\\
\hline
\enddata
\end{deluxetable}

%% manuscript produces a one-column, double-spaced document:

%% preprint2 produces a double-column, single-spaced document:

\begin{deluxetable}{|c|c|c|r|r|r|r|r|r|r|r|r|r|r|r|r|r|r|r|r|r|r|r|}
\tabletypesize{\tiny}
\rotate
\tablecaption{2MASS/IRAS/MSX PHOTOMETRY FOR DIRBE CATALOG SOURCES \label{tbl-4}}
\tablewidth{0pt}
\tablehead{
\colhead{DIRBE name}&
\colhead{TWOMASS name}&
\colhead{IRAS name}&
\multicolumn{6}{c}{2MASS PHOTOMETRY}&
\multicolumn{4}{c}{IRAS PHOTOMETRY}&
\\
\colhead{}&
\colhead{}&
\colhead{}&
\colhead{J }&
\colhead{$\sigma$$_J$}&
\colhead{H}&
\colhead{$\sigma$$_H$}&
\colhead{K}&
\colhead{$\sigma$$_K$}&
\colhead{F$_{12}$}&
\colhead{F$_{25}$}&
\colhead{F$_{60}$}&
\colhead{F$_{100}$}
\\
\colhead{}&
\colhead{}&
\colhead{}&
\colhead{(mag)}&
\colhead{(mag)}&
\colhead{(mag)}&
\colhead{(mag)}&
\colhead{(mag)}&
\colhead{(mag)}&
\colhead{(Jy)}&
\colhead{(Jy)}&
\colhead{(Jy)}&
\colhead{(Jy)}
\\
}
\startdata
D00000657P2553112 & 00000657+2553112 & IRAS23575+2536  &          2.225 &  0.296&   1.317 &  0.158 &  0.915 &  0.188 &   53.780 &   22.560  &   3.330  &   1.783\\ 
D00000690P2014145 & 00000690+2014145 &&                          4.026 &  0.228 &   3.079 &   0.206 &  2.562 &   0.276
&-99.990
&-99.990
&-99.990
&-99.990
\\
D00001353P5541579 & 00001353+5541579   &&                        4.490  & 0.292  & 3.479   & 0.216  & 2.987   & 0.288   
&-99.990
&-99.990
&-99.990
&-99.990
\\
D00001815P6021016 & 00001815+6021016     &&                      3.557   & 0.258 &  2.640  &  0.204 &  2.182  &  0.240 
&-99.990
&-99.990
&-99.990
&-99.990
\\
D00010244P3830145 & 00010244+3830145 &IRAS23584+3813  &          3.542   & 0.248 &  2.630 &  0.178   & 2.083   & 0.234 &
19.770 &    8.800 &    1.410 &    1.011

 \\
\enddata
\tablenotetext{a}{NOTE:-Table 3 is presented in its entirety, 
including additional columns with IRAS flags and VAR parameter,
in the electronic
edition of the Astrophysical Journal Supplement.  A portion is shown here for guidance regarding
its form and content. }
\end{deluxetable}

%% manuscript produces a one-column, double-spaced document:

%% preprint2 produces a double-column, single-spaced document:

\begin{deluxetable}{crcrr}
\tabletypesize{\scriptsize}
\tablecaption{COMPANION INFORMATION\label{tbl-5}}
\tablewidth{0pt}
\tablehead{
\colhead{DIRBE}&
\colhead{WAVELENGTH}&
\colhead{COMPANION}&
\colhead{F$_{\nu}$}&
\colhead{OFFSET}
\\
\colhead{NAME} & 
\colhead{($\mu$m)} & 
\colhead{NAME} & 
\colhead{(Jy)}   & \colhead{(degrees)} \\
}
\startdata
D00001353P5541579 &  1.25 & 2MASS00032146+5540518 &       124.1 & 0.441882 \\
D00001353P5541579 &  2.20 & 2MASS00032146+5540518 &       234.4 & 0.441882 \\
D00001353P5541579 &  3.50 & AFGL 5                &       199.2 & 0.435020 \\
D00001353P5541579 &  4.90 & AFGL 5                &       186.4 & 0.435020 \\
D00001353P5541579 &  4.90 & RAFGL 5               &       155.0 & 0.435020 \\
D00001353P5541579 & 12.00 & IRAS00007+5524        &        97.6 & 0.441621 \\
\enddata
\tablenotetext{a}{NOTE:-Table 4 is presented in its entirety
in the electronic
edition of the Astronomical Journal.  A portion is shown here for guidance regarding
its form and content. }
\end{deluxetable}

\begin{deluxetable}{|c|r|r|r|r|r|r|}
\tabletypesize{\scriptsize}
%\tabletypesize{\tiny}
\rotate
\tablecaption{INCOMPLETENESS OF DIRBE CATALOG DUE TO CONFUSION\label{tbl-6}}
\tablewidth{0pt}
\tablehead{
\colhead{}&
\colhead{1.25 $\mu$m}&   
\colhead{2.2 $\mu$m}&   
\colhead{12 $\mu$m}&   
\colhead{25 $\mu$m}&   
\colhead{60 $\mu$m}&   
\colhead{100 $\mu$m}
\\
}
\startdata
 Nominal Completeness Limits (Jy) &  90  &  60 & 90 & 165 & 960 & 2295 \\ 
 Number in IRAS/2MASS/MSX $>$ DIRBE 3$\sigma$ limit &   2511 &   5389 & 
  727 &   485 &   461 &   454 \\ 
 Number of these with all points removed by filtering &   685 &   1211 & 
  163 &   262 &   336 &   392 \\ 
 Percent with all removed by filtering &  27 &  22 &  22 &  54 &  72 &  86 \\ 
\enddata
\end{deluxetable}

\begin{deluxetable}{|c|r|r|r|r|r|r|r|r|r|r|}
%\tabletypesize{\scriptsize}
\tabletypesize{\tiny}
\rotate
\tablecaption{STATISTICS ON CATALOG SUBSET WITH 2MASS K $<$ 1.0\label{tbl-7}}
\tablewidth{0pt}
\tablehead{
\colhead{}&
\colhead{1.25 $\mu$m}&   
\colhead{2.2 $\mu$m}&   
\colhead{3.5 $\mu$m}&   
\colhead{4.9 $\mu$m}&   
\colhead{12 $\mu$m}&   
\colhead{25 $\mu$m}&   
\colhead{60 $\mu$m}&   
\colhead{100 $\mu$m}&   
\colhead{140 $\mu$m}&   
\colhead{240 $\mu$m}
\\
}
\startdata
Number Detected &
&&&&&&&&&\\
with S/N $>$ 3&
462&570&765&699&279&98&14&14&1&2\\
\hline
Number Flagged for &&&&&&&&&&\\
Companion in Beam&153&252&20&138&120&52&22&18&0&0\\
\hline
Number Flagged for &&&&&&&&&&\\
Discrepancy with Other Photometry&4&1&0&0&63&18&7&12&0&0\\
\hline
Number Flagged for&&&&&&&&&&\\
2-week rms/$<$error$>$ $\ge$ 3&45&92&213&98&106&89&139&165&62&63\\
\hline
Number Unflagged $\ge$3$\sigma$
&&&&&&&&&&\\
Detections with N $\ge$ 100&
215&271&540&435&149&52&5&1&1&1\\
\hline
Number Unflagged $>$3$\sigma$ Detections&&&&&&&&&&\\
with $>$3$\sigma$ Amplitudes and N $\ge$ 100
&57&90&123&77&20&11&0&0&0&0\\
\hline
\enddata
\end{deluxetable}

%% manuscript produces a one-column, double-spaced document:

%% preprint2 produces a double-column, single-spaced document:

\begin{deluxetable}{|c|c|c|r|r|r|r|r|r|r|r|r|r|r|r|}
\tabletypesize{\tiny}
\rotate
\tablecaption{DIRBE CATALOG ASSOCIATIONS AND OTHER INFORMATION\label{tbl-8}}
\tablewidth{0pt}
\tablehead{
\colhead{DIRBE NAME}&
\colhead{TWOMASS NAME}&
\colhead{IRAS NAME}&
\colhead{FIRST}&
\colhead{SIMBAD}&
\colhead{SIMBAD}&
\colhead{SIMBAD}&
\colhead{SECOND}&
\\
\colhead{}&
\colhead{}&
\colhead{}&
\colhead{SIMBAD}&
\colhead{OBJECT}&
\colhead{SPECTRAL}&
\colhead{OFFSET}&
\colhead{SIMBAD}
\\
\colhead{}&
\colhead{}&
\colhead{}&
\colhead{NAME}&
\colhead{TYPE}&
\colhead{TYPE}&
\colhead{($''$)}&
\colhead{NAME}
\\
}
\startdata
D00000657P2553112  & 00000657+2553112  & IRAS23575+2536  &        V* Z Peg      &             Mi*  & M7e   &       0.2 & \\      
D00000690P2014145  &  00000690+2014145  &                          &  V* EP Peg                   &  sr*  &  M
...         &   0.1  & IRAS 23575+1957           \\
D00001353P5541579  &  00001353+5541579  &                          &  HD 224719                   &    *  &  M
...         &   0.1  &                           \\
D00001815P6021016  &  00001815+6021016  &                          &  HD 224754                   &    *  &  M2           &   0.1  &                           \\
D00010244P3830145  &  00010244+3830145  &  IRAS23584+3813          &  DO 22623                    &    *  &  M0           &   5.4  &                           \\
\enddata
\tablenotetext{a}{NOTE:-Table 7 is presented in its entirety, 
including additional columns with more SIMBAD data,
GCVS/NSV/NSV information, 
IRAS spectral types,
and OH/IR star 
identifications, in the electronic
edition of the Astrophysical Journal Supplement.  
A portion is shown here for guidance regarding
its form and content. }
\end{deluxetable}

\begin{deluxetable}{|c|r|r|r|r|r|r|r|r|r|r|r|}
%\tabletypesize{\scriptsize}
\tabletypesize{\tiny}
\tablecaption{STATISTICS ON ASSOCIATIONS: FULL CATALOG\label{tbl-9}}
\tablewidth{0pt}
\tablehead{
\colhead{Type}&
\colhead{}&
\colhead{1.25 $\mu$m}&   
\colhead{2.2 $\mu$m}&   
\colhead{3.5 $\mu$m}&   
\colhead{4.9 $\mu$m}&   
\colhead{12 $\mu$m}&   
\colhead{25 $\mu$m}&   
\colhead{60 $\mu$m}&   
\colhead{100 $\mu$m}&   
\colhead{140 $\mu$m}&   
\colhead{240 $\mu$m}
\\
}
\startdata
\hline
  M Mira/SRa/SRb/Lb& S/N $\ge$ 3 per datapoint     &  1353  &  1802  &  1732  &  1433  &   249  &    98  &    18  &    20  &     1  &     1 \\ 
                   & unflagged with N $\ge$ 100  &    612 &    836 &   1157 &    875 &    132 &     51 &      5 &      3 &      1 &      1 \\ 
                   & $\ge$3$\sigma$ amplitude  &     94 &    160 &    191 &    127 &     30 &     17 &      0 &      0 &      0 &      0 \\ 
                   & percent variable &     15 &     19 &     17 &     15 &     23 &     33 &      0 &      0 &      0 &      0 \\ 
M other/unspecified& S/N $\ge$ 3 per datapoint     &  1394  &  1905  &  1468  &  1031  &   113  &    54  &    41  &    51  &     8  &     7 \\ 
                   & unflagged with N $\ge$ 100  &    517 &    755 &    834 &    524 &     41 &     14 &      5 &     12 &      2 &      2 \\ 
                   & $\ge$3$\sigma$ amplitude  &      6 &     13 &     29 &     13 &     10 &      3 &      0 &      1 &      0 &      0 \\ 
                   & percent variable &      1 &      2 &      3 &      2 &     24 &     21 &      0 &      8 &      0 &      0 \\
\enddata
\end{deluxetable}

\begin{deluxetable}{|c|r|r|r|r|r|r|r|r|r|r|r|}
%\tabletypesize{\scriptsize}
\tabletypesize{\tiny}
\tablecaption{STATISTICS ON ASSOCIATIONS: 2MASS K $<$ 1.0 SUBSAMPLE \label{tbl-10}}
\tablewidth{0pt}
\tablehead{
\colhead{Type}&
\colhead{}&
\colhead{1.25 $\mu$m}&   
\colhead{2.2 $\mu$m}&   
\colhead{3.5 $\mu$m}&   
\colhead{4.9 $\mu$m}&   
\colhead{12 $\mu$m}&   
\colhead{25 $\mu$m}&   
\colhead{60 $\mu$m}&   
\colhead{100 $\mu$m}&   
\colhead{140 $\mu$m}&   
\colhead{240 $\mu$m}
\\
}
\startdata
\hline
  M Mira/SRa/SRb/Lb& S/N $\ge$ 3 per datapoint     &   266  &   316  &   417  &   387  &   173  &    62  &     8  &     8  &     0  &     0 \\ 
                   & unflagged with N $\ge$ 100  &    129 &    161 &    302 &    251 &     98 &     34 &      4 &      1 &      0  &      0 \\ 
                   & $\ge$3$\sigma$ amplitude  &     41 &     69 &     91 &     61 &     11 &      6 &      0 &      0 &      0 &      0 \\ 
                   & percent variable &     32 &     43 &     30 &     24 &     11 &     18 &      0 &      0 &     $-$ &     $-$ \\ 
M other/unspecified& S/N $\ge$ 3 per datapoint     &    76  &    98  &   138  &   121  &    39  &    17  &     4  &     2  &     1  &     1 \\ 
                   & unflagged with N $\ge$ 100  &     35 &     42 &     93 &     72  &     15  &      7  &      1  &      0  &      1  &      1  \\ 
                   & $\ge$3$\sigma$ amplitude  &      3  &      7  &     11  &      4  &      1  &      1  &      0  &      0  &      0  &      0  \\ 
                   & percent variable &      9  &      17 &     12  &      6  &      7  &      14  &      0  &      $-$  &     0 &     0 \\
\enddata
\end{deluxetable}

%% manuscript produces a one-column, double-spaced document:

%% preprint2 produces a double-column, single-spaced document:

\begin{deluxetable}{|c|c|c|r|r|r|r|r|r|c|c|c|c|c|c|}
\tabletypesize{\scriptsize}
\rotate
\tablecaption{BRIGHTEST 1.25 $\mu$m SOURCES IN DIRBE CATALOG\label{tbl-11}}
\tablewidth{0pt}
\tablehead{
\colhead{DIRBE}&
\colhead{2MASS}&
\colhead{IRAS}&
\multicolumn{7}{c}{1.25 $\mu$m}&
\multicolumn{1}{c}{SIMBAD}&
\multicolumn{1}{c}{SPECTRAL}&
\multicolumn{1}{c}{VAR}
\\
\colhead{NAME} & 
\colhead{NAME}&
\colhead{NAME}&
\colhead{F$_{\nu}$}   & \colhead{$\sigma$}   &
\colhead{$<$$err$$>$}&
\colhead{$\Delta$mag} &
\colhead{$\sigma$($\Delta$mag)} &
\colhead{N}&
\colhead{Flags$^a$}&
\colhead{NAME}   & 
\colhead{TYPE}   &
\colhead{TYPE}   &
\\
}
\startdata
D05551028P0724255 & 05551028+0724255 & 05524+0723  & 30122 &    922 &     510 &  0.08 &  0.03 &   99 &  
000 & $\alpha$ Ori & M1 & SR \\
D04364544M6204379 & 04364544-6204379 & 04361-6210 &  16977 &    642 &    386 &  0.02 &  0.04 &  26 &
000 &  R Dor    &  M8III & SR\\
D17143885P1423253 & 17143885+1423253 & 17123+1426         &                      13590 &    362 &     388 & 0.02 &  0.04  & 26 &
000  & $\alpha$ Her & M5Ib-II & SR \\
D14153968P1910558 & 14153968+1910558 & 14133+1925 &                             12763 &     247 &     248 &  0.04 &   0.03 &  229 &
000 & $\alpha$ Boo   &        K1.5III &                    \\
D22424003M4653044 & 22424003-4653044 & 22396-4708
                            &  11397 &    398 &    234 &  0.06 &  0.03 & 123 &
001 & $\beta$ Gru   & M5III &   \\                   
D13490199M2822034 & 13490199-2822034 & 13462-2807
                 &               7331 &   1753 &    123 &  0.11 &  0.02 &  29 &
000 &  W Hya &  M7e & SR \\
D05164138P4559525 &  05164138+4559525 & 05130+4556 &                            5451 &     72 &     105 &   0.01  &  0.03 &  61 &  
000 &
$\alpha$ Aur &  G5IIIe+...  &   \\    
D23034644P2804580 & 13490199-2822034 & 13462-2807     &                          4781 &    109 &     93 &  0.05 &  0.03 &  297 &
000 &  $\beta$ Peg     & M2.5II-III                 & \\
\enddata
\tablenotetext{a}{If the first confusion flag is set to 1, there is 
another source in the DIRBE beam with a 2MASS J band flux density above 25 Jy.
If the second confusion flag is set to 1, the DIRBE photometry at minimum (in the weekly-averaged light curve) is greater 
than that of 2MASS by more than 3$\sigma$, or the DIRBE photometry
at maximum is less 
than that of 2MASS by more than 3$\sigma$.
If this flag is set, this may mean a second source
in the beam, extended emission, or large variations not observed in DIRBE.
If the third flag is set, the rms/$<$errorbar$>$ for a 2-week period is greater than
3, suggesting confusion from an unknown source that is not completely corrected by
filtering.
}
\end{deluxetable}

%% manuscript produces a one-column, double-spaced document:

%% preprint2 produces a double-column, single-spaced document:

\begin{deluxetable}{|c|c|c|r|r|r|r|r|r|c|c|c|c|c|c|}
\tabletypesize{\scriptsize}
\rotate
\tablecaption{BRIGHTEST 2.2 $\mu$m SOURCES IN DIRBE CATALOG\label{tbl-12}}
\tablewidth{0pt}
\tablehead{
\colhead{DIRBE}&
\colhead{2MASS}&
\colhead{IRAS}&
\multicolumn{7}{c}{2.2 $\mu$m}&
\multicolumn{1}{c}{SIMBAD}&
\multicolumn{1}{c}{SPECTRAL}&
\multicolumn{1}{c}{VAR}
\\
\colhead{NAME} & 
\colhead{NAME} & 
\colhead{NAME} & 
\colhead{F$_{\nu}$}   & \colhead{$\sigma$}   &
\colhead{$<$$err$$>$}&
\colhead{$\Delta$mag} &
\colhead{$\sigma$($\Delta$mag)} &
\colhead{N}&
\colhead{Flags$^a$}&
\colhead{NAME}   & 
\colhead{TYPE}   &
\colhead{TYPE}   &
\\
}
\startdata
D05551028P0724255 & 05551028+0724255 & 05524+0723          & 29995 &  889 &  210 & 0.07 & 0.01 &   72 & 100 &    $\alpha$ Ori                                           &  M1 & SR                         \\
D04364544M6204379 & 04364544-6204379 & 04361-6210          & 25601 & 1898 &  163 & 0.19 & 0.01 &  119 & 000 &    R Dor                                             &  M8IIIe & SR                    \\
D16292443M2625549 & 16292443-2625549 & 16262-2619A         & 23270 &  179 &  143 & 0.01 & 0.01 &   44 & 100 &    $\alpha$ Sco                &  M1.5Ib   & SR   \\
D17143885P1423253 & 17143885+1423253 & 17123+1426          & 15614 &  229 &  107 & 0.04 & 0.01 &  193 & 000 &    $\alpha$ Her       &   M5II  &  SR     \\
D22424003M4653044 & 22424003-4653044 & 22396-4708          & 12123 &  107 &   78 & 0.02 & 0.01 &  190 & 000 &    $\beta$ Gru                                           &  M5III   &                   \\
D12310993M5706474 & 12310993-5706474 & 12283-5650          & 11710 &   82 &   97 & 0.00 & 0.01 &   35 & 100 &    $\gamma$ Cru                                            &   M3.5III &                   \\
D13490199M2822034 & 13490199-2822034 & 13462-2807          & 11442 & 2506 &   85 & 0.50 & 0.01 &  160 & 000 &    W Hya &  M7e            & SR   \\
D14153968P1910558 & 14153968+1910558 & 14133+1925          & 10073 &   61 &   62 & 0.01 & 0.01 &  299 & 000 &    $\alpha$ Boo                                          &  K1.5III  &                  \\
\enddata
\tablenotetext{a}{If the first confusion flag is set to 1, there is 
another source in the DIRBE beam with a 2MASS K band flux density above 20 Jy.
If the second confusion flag is set to 1, the DIRBE photometry at minimum in the weekly-averaged light curve is greater 
than that of 2MASS by more than 3$\sigma$, or the DIRBE photometry
at maximum is less 
than that of 2MASS by more than 3$\sigma$.
If this flag is set, this may mean a second source
in the beam, extended emission, or large variations not observed in DIRBE.
If the third flag is set, the rms/$<$errorbar$>$ for a 2-week period is greater than
3, suggesting confusion from an unknown source that is not completely corrected by
filtering.
}
\end{deluxetable}

%% manuscript produces a one-column, double-spaced document:

%% preprint2 produces a double-column, single-spaced document:

\begin{deluxetable}{|c|c|c|r|r|r|r|r|r|c|c|c|c|c|c|}
\tabletypesize{\scriptsize}
\rotate
\tablecaption{BRIGHTEST 3.5 $\mu$m SOURCES IN DIRBE CATALOG\label{tbl-13}}
\tablewidth{0pt}
\tablehead{
\colhead{DIRBE}&
\colhead{2MASS}&
\colhead{IRAS}&
\multicolumn{7}{c}{3.5 $\mu$m}&
\multicolumn{1}{c}{SIMBAD}&
\multicolumn{1}{c}{SPECTRAL}&
\multicolumn{1}{c}{VAR}
\\
\colhead{NAME} & 
\colhead{NAME} & 
\colhead{NAME} & 
\colhead{F$_{\nu}$}   & \colhead{$\sigma$}   &
\colhead{$<$$err$$>$}&
\colhead{$\Delta$mag} &
\colhead{$\sigma$($\Delta$mag)} &
\colhead{N}&
\colhead{Flags$^a$}&
\colhead{NAME}   & 
\colhead{TYPE}   &
\colhead{TYPE}   &
\\
}
\startdata
D05551028P0724255 & 05551028+0724255 & 05524+0723          & 17065 &  546 &  228 & 0.08 & 0.02 &  312 & 000 &    $\alpha$ Ori                                           &  M1 & SR                         \\
D04364544M6204379 & 04364544-6204379 & 04361-6210          & 15611 &  955 &  233 & 0.21 & 0.02 &  684 & 000 &    R Dor                                             &  M8IIIe  & SR                   \\
D16292443M2625549 & 16292443-2625549 & 16262-2619A         & 13063 &  233 &  188 & 0.05 & 0.02 &  274 & 000 &     $\alpha$ Sco                              &  M1.5Ib & SR   \\
D17143885P1423253 & 17143885+1423253 & 17123+1426          &  8310 &  160 &  122 & 0.06 & 0.02 &  475 & 000 &     $\alpha$ Her                                 &  M5II  & SR    \\
D13490199M2822034 & 13490199-2822034 & 13462-2807          &  7945 & 1107 &  105 & 0.34 & 0.02 &  366 & 000 &    W Hya  &  M7e & SR              \\
D22424003M4653044 & 22424003-4653044 & 22396-4708          &  6248 &  124 &   89 & 0.04 & 0.02 &  317 & 001 &    $\beta$ Gru                                           &  M5III  &                    \\
D12310993M5706474 & 12310993-5706474 & 12283-5650          &  5920 &   90 &   89 & 0.03 & 0.02 &  803 & 000 &    $\gamma$ Cru                                            &   M3.5III &                    \\
D09473348P1125436 & 09473348+1125436 & 09448+1139          &  5104 &  589 &   72 & 0.33 & 0.02 &  179 & 000 &    R Leo     &  M8IIIe  & M            \\
\enddata
\tablenotetext{a}{If the first confusion flag is 
set to 1, there is 
another source in the DIRBE beam with a 3.5 $\mu$m flux density above 20 Jy in the
Catalog of Infrared Observations.
The second confusion flag is not set at 3.5 $\mu$m.
If the third flag is set, the rms/$<$errorbar$>$ for a 2-week period is greater than
3, suggesting confusion from an unknown source that is not completely corrected by
filtering.
}
\end{deluxetable}

%% manuscript produces a one-column, double-spaced document:

%% preprint2 produces a double-column, single-spaced document:

\begin{deluxetable}{|c|c|c|r|r|r|r|r|r|c|c|c|c|c|c|}
\tabletypesize{\scriptsize}
\rotate
\tablecaption{BRIGHTEST 4.9 $\mu$m SOURCES IN DIRBE CATALOG\label{tbl-14}}
\tablewidth{0pt}
\tablehead{
\colhead{DIRBE}&
\colhead{2MASS}&
\colhead{IRAS}&
\multicolumn{7}{c}{4.9 $\mu$m}&
\multicolumn{1}{c}{SIMBAD}&
\multicolumn{1}{c}{SPECTRAL}&
\multicolumn{1}{c}{VAR}
\\
\colhead{NAME} & 
\colhead{NAME} & 
\colhead{NAME} & 
\colhead{F$_{\nu}$}   & \colhead{$\sigma$}   &
\colhead{$<$$err$$>$}&
\colhead{$\Delta$mag} &
\colhead{$\sigma$($\Delta$mag)} &
\colhead{N}&
\colhead{Flags$^a$}&
\colhead{NAME}   & 
\colhead{TYPE}   &
\colhead{TYPE}   &
\\
}
\startdata
D09475740P1316435 & 09475740+1316435 & 09452+1330 & 9011  &   5742 &      61 &  1.37 &  0.01 &  89 &  
001  & CW Leo  & C & Mira                           \\
D04364544M6204379  & 04364544-6204379 & 04361-6210  &    8729 &   256 &   41 &   0.11 &  0.01 & 530 &
000 & R Dor & M8IIIe & SR \\
D05551028P0724255 & 05551028+0724255 & 05524+0723  & 7142 &   326 &     31 &   0.11 &  0.01 &   67 &
000 & $\alpha$ Ori & M1 & SR \\
D16292443M2625549 & 16292443-2625549 & 16262-2619A &  5650 &   31 &   25 &  0.01 &   0.01 &  71 &
100 & $\alpha$ Sco & M1Ib+ & SR \\
D13490199M2822034 & 13490199-2822034 & 13462-2807 & 5378 &  326 &   25 &  0.18 &  0.01 &  297 &
001 & W Hya & M7e & M \\
D17143885P1423253 & 17143885+1423253 & 17123+1426 & 3616 &   54 & 17 &  0.04 &  0.01 & 135 &
100 & $\alpha$ Her & M5II & SR \\
D09473348P1125436  & 09473348+1125436 & 09448+1139 & 3459  & 364 &  17 &  0.28 &  0.01 &  91 &
101 & R Leo & M8IIIe & M \\
D02192081M0258393 & 02192081-0258393 & 02168-0312 & 3221 &  812 &  17 &  0.70 &  0.01 &  148 &
001 & omi Cet & M7IIIe & M \\
\enddata
\tablenotetext{a}{If the first confusion flag is set to 1, there is 
another source in the DIRBE beam with a 4.9 $\mu$m flux density above 10 Jy in
the Catalog of Infrared Observations or in the Egan $\&$ Price (1996) synthetic 4.2 $\mu$m catalog.
The second confusion flag is not set at 4.9 $\mu$m.
If the third flag is set, the rms/$<$errorbar$>$ for a 2-week period is greater than
3, suggesting confusion from an unknown source that is not completely corrected by
filtering.
}
\end{deluxetable}

%% manuscript produces a one-column, double-spaced document:

%% preprint2 produces a double-column, single-spaced document:

\begin{deluxetable}{|c|c|c|r|r|r|r|r|r|c|c|c|c|c|c|}
\tabletypesize{\scriptsize}
\rotate
\tablecaption{BRIGHTEST 12 $\mu$m SOURCES IN DIRBE CATALOG\label{tbl-15}}
\tablewidth{0pt}
\tablehead{
\colhead{DIRBE}&
\colhead{2MASS}&
\colhead{IRAS}&
\multicolumn{7}{c}{12 $\mu$m}&
\multicolumn{1}{c}{SIMBAD}&
\multicolumn{1}{c}{SPECTRAL}&
\multicolumn{1}{c}{VAR}
\\
\colhead{NAME} & 
\colhead{NAME} & 
\colhead{NAME} & 
\colhead{F$_{\nu}$}   & \colhead{$\sigma$}   &
\colhead{$<$$err$$>$}&
\colhead{$\Delta$mag} &
\colhead{$\sigma$($\Delta$mag)} &
\colhead{N}&
\colhead{Flags$^a$}&
\colhead{NAME}   & 
\colhead{TYPE}   &
\colhead{TYPE}   &
\\
}
\startdata
D09475740P1316435 & 09475740+1316435 & 09452+1330          & 31773 & 7661 &  356 & 0.33 & 0.02 &   51 & 110 &    CW Leo                                            & C & M                      \\
D05351646M0523230 & 05351646-0523230 &  & 19690 &  491 &  266 & 0.03 & 0.02 &   59 & 100 &                     $\theta$ Ori C                &                O6pe   &    \\
D05353020M0515490 & & 05330-0517          & 19114 & 1205 &  254 & 0.08 & 0.02 &   74 & 111 &                                              &                         &      \\
D05351000M0527470 & & 05327-0529          & 18596 & 1321 &  284 & 0.05 & 0.02 &   48 & 110 &     LP Ori                 &                 B1.5Vp & Orion     \\
D05353430M0514520 &  & 05331-0515          & 17565 & 2292 &  253 & 0.09 & 0.02 &   35 & 111 &                                           &                          &     \\
D11142900M6118000 &  & 11123-6101          &  9518 &  847 &  166 & 0.16 & 0.03 &   82 & 111 & OH CSI-61-11124                                         & WR        &                 \\
D07225830M2546030 & 07225830-2546030 & 07209-2540          &  9278 &  299 &   88 & 0.07 & 0.01 &   66 & 110 &   VY CMa                  &  M3/M4II:               & \\
D11155060M6113420 &  & 11136-6056          &  8676 &  516 &  136 & 0.10 & 0.02 &   78 & 111 &                                       &                             &  \\
\enddata
\tablenotetext{a}{If the first confusion flag is set to 1, there is 
another source in the DIRBE beam with an IRAS 12 $\mu$m flux 
density above 30 Jy.
If the second confusion flag is set to 1, the DIRBE photometry at minimum is greater 
than that of IRAS by more than 3$\sigma$, or the DIRBE photometry
at maximum is less 
than that of IRAS by more than 3$\sigma$.
If this flag is set, this may mean a second source
in the beam, extended emission, or large variations not observed in DIRBE.
If the third flag is set, the rms/$<$errorbar$>$ for a 2-week period is greater than
3, suggesting confusion from an unknown source that is not completely corrected by
filtering.
}
\end{deluxetable}

%% manuscript produces a one-column, double-spaced document:

%% preprint2 produces a double-column, single-spaced document:

\begin{deluxetable}{|c|c|c|r|r|r|r|r|r|c|c|c|c|c|c|}
\tabletypesize{\scriptsize}
\rotate
\tablecaption{BRIGHTEST 25 $\mu$m SOURCES IN DIRBE CATALOG\label{tbl-16}}
\tablewidth{0pt}
\tablehead{
\colhead{DIRBE}&
\colhead{2MASS}&
\colhead{IRAS}&
\multicolumn{7}{c}{25 $\mu$m}&
\multicolumn{1}{c}{SIMBAD}&
\multicolumn{1}{c}{SPECTRAL}&
\multicolumn{1}{c}{VAR}
\\
\colhead{NAME} & 
\colhead{NAME} & 
\colhead{NAME} & 
\colhead{F$_{\nu}$}   & \colhead{$\sigma$}   &
\colhead{$<$$err$$>$}&
\colhead{$\Delta$mag} &
\colhead{$\sigma$($\Delta$mag)} &
\colhead{N}&
\colhead{Flags$^a$}&
\colhead{NAME}   & 
\colhead{TYPE}   &
\colhead{TYPE}   &
\\
}
\startdata
D08590440M4730370&&08573-4718&                                  
28682 &    951 &     200 &   0.00 &  0.01 &  24 & 
111 & Maser 267.94-01.06   &                          &\\

D09475740P1316435 & 09475740+1316435 & 09452+1330          & 28611 & 4508 &  230 & 0.18 & 0.01 &   47 & 110 &    CW Leo                                            &  C & M                      \\
D10463100M6003470 &  & 10445-5947          & 17092 & 3716.&  642 & 0.01 & 0.06 &   23 & 110 &&                                       &                              \\
D07225830M2546030 & 07225830-2546030 & 07209-2540          & 12346 &  345 &   96. & 0.10 & 0.01 &  397 & 110 &  VY CMa    &  M3/M4II:              &   \\
D18172538M1145416&
18172538-1145416 
& 18146-1146 & 10368 &   2143 &     677 &   0.06  &  0.10 &  26 & 110 &&&\\
D10482740M5951210 &  & 10464-5935          &  8540 & 4059 &  829 & 0.28 & 0.16 &   32 & 110  &                                         &            &                  \\
D20343600P4105541 & 20343600+4105541 &  &  6394 &  591 &  806 & 0.04 & 0.21 &   36 & 100 &    HD 196241                                            &  K5   &                      \\
D18312470M0205320                                 
&
& 18288-0207
&
6162 &   229 &    79 &   0.03 &  0.02 & 46 &111& W40&& \\
\enddata
\tablenotetext{a}{If the first confusion flag is set to 1, there is 
another source in the DIRBE beam with an IRAS
25 $\mu$m flux density above 55 Jy.
If the second confusion flag is set to 1, the DIRBE photometry at minimum is greater 
than that of IRAS by more than 3$\sigma$, or the DIRBE photometry
at maximum is less 
than that of IRAS by more than 3$\sigma$.
If this flag is set, this may mean a second source
in the beam, extended emission, or large variations not observed in DIRBE.
If the third flag is set, the rms/$<$errorbar$>$ for a 2-week period is greater than
3, suggesting confusion from an unknown source that is not completely corrected by
filtering.
}
\end{deluxetable}

\begin{deluxetable}{c|c|ccc|ccc|ccc|ccc|ccc|ccc|ccc|ccc}
%\tabletypesize{\scriptsize}
\rotate
\tabletypesize{\tiny}
\tablecaption{MEAN DIRBE AMPLITUDES OF VARIATION\label{tbl-17}}
\tablewidth{0pt}
\tablehead{
\colhead{Var}&
\colhead{Spectral} & 
\multicolumn{3}{c}{1.25 {$\mu$}m }&
\multicolumn{3}{c}{2.2 {$\mu$}m }&
\multicolumn{3}{c}{3.5 {$\mu$}m }&
\multicolumn{3}{c}{4.9 {$\mu$}m }&
\multicolumn{3}{c}{12 {$\mu$}m }&
\multicolumn{3}{c}{25 {$\mu$}m }
\\
\colhead{Type}&
\colhead{Type} & 
\colhead{$<$${\Delta}$mag$>$}   &
\colhead{$<$${\sigma}$$>$}   &
\colhead{N} &
\colhead{$<$${\Delta}$mag$>$}   &
\colhead{$<$${\sigma}$$>$}   &
\colhead{N} &
\colhead{$<$${\Delta}$mag$>$}   &
\colhead{$<$${\sigma}$$>$}   &
\colhead{N} &
\colhead{$<$${\Delta}$mag$>$}   &
\colhead{$<$${\sigma}$$>$}   &
\colhead{N} &
\colhead{$<$${\Delta}$mag$>$}   &
\colhead{$<$${\sigma}$$>$}   &
\colhead{N} &
\colhead{$<$${\Delta}$mag$>$}   &
\colhead{$<$${\sigma}$$>$}   &
\colhead{N} 
\\
}
\startdata
      Mira &       OH/IR &  0.82$\pm$0.58&  0.15&        12 &  0.58$\pm$0.31&  0.13&        36 &  0.68$\pm$0.29&  0.11&        31 &  0.55$\pm$0.28&  0.13&        37 &  0.66$\pm$0.23&  0.13&        13 &  0.61$\pm$0.16&  0.12&         9\\
      Mira &      oxygen &  0.53$\pm$0.33&  0.11&        93 &  0.45$\pm$0.25&  0.12&       160 &  0.50$\pm$0.23&  0.12&       152 &  0.37$\pm$0.21&  0.13&       174 &  0.60$\pm$0.27&  0.15&        26 &  0.64$\pm$0.22&  0.18&        12\\
      Mira &      carbon &  0.56$\pm$0.26&  0.18&         4 &  0.64$\pm$0.42&  0.12&        17 &  0.77$\pm$0.35&  0.11&        20 &  0.45$\pm$0.35&  0.10&        38 &  0.59$\pm$0.18&  0.14&        18 &  0.52$\pm$0.19&  0.14&         7\\
      Mira &           S &  0.46$\pm$0.27&  0.13&         3 &  0.34$\pm$0.23&  0.14&         7 &  0.67$\pm$0.23&  0.11&         9 &  0.38$\pm$0.19&  0.14&         9 &  0.92     &  0.10&         1 &  0.88     &  0.23&         1\\
       SRa &       OH/IR &    $-$    &  0.13&         0 &  0.74     &  0.24&         1 &    $-$    &  0.11&         0 &    $-$    &  0.14&         0 &  0.61     &  0.09&         1 &  0.70     &  0.22&         1\\
       SRa &      oxygen &  0.22$\pm$0.21&  0.13&        39 &  0.19$\pm$0.13&  0.12&        46 &  0.26$\pm$0.16&  0.12&        36 &  0.17$\pm$0.12&  0.12&        30 &  0.22$\pm$0.09&  0.15&         4 &  0.22     &  0.17&         1\\
       SRa &      carbon &  0.45$\pm$0.41&  0.10&         5 &  0.24$\pm$0.20&  0.10&         5 &  0.30$\pm$0.16&  0.10&         7 &  0.29$\pm$0.17&  0.11&         5 &  0.16     &  0.03&         1 &    $-$    &  0.17&         0\\
       SRb &       OH/IR &    $-$    &  0.19&         0 &  0.24     &  0.17&         1 &  0.35     &  0.24&         1 &    $-$    &  0.11&         0 &    $-$    &  0.03&         0 &    $-$    &  0.17&         0\\
\enddata
\end{deluxetable}

\begin{deluxetable}{c|c|cc|cc|cc|cc|cc|cc|cc|cc}
%\tabletypesize{\scriptsize}
\tabletypesize{\tiny}
\tablecaption{MEAN AMPLITUDE RATIOS\label{tbl-18}}
\tablewidth{0pt}
\tablehead{
\colhead{Variability}&
\colhead{Spectral} & 
\multicolumn{2}{c}{1.25/2.2}&
\multicolumn{2}{c}{2.2/3.5 }&
\multicolumn{2}{c}{3.5/4.9 }&
\multicolumn{2}{c}{4.9/12}&
\multicolumn{2}{c}{12/25}
\\
\colhead{Type}&
\colhead{Type} & 
\colhead{${\Delta}$mag/${\Delta}$mag}   &
\colhead{N} &
\colhead{${\Delta}$mag/${\Delta}$mag}   &
\colhead{N} &
\colhead{${\Delta}$mag/${\Delta}$mag}   &
\colhead{N} &
\colhead{${\Delta}$mag/${\Delta}$mag}   &
\colhead{N} &
\colhead{${\Delta}$mag/${\Delta}$mag}   &
\colhead{N} 
\\
}
\startdata
      Mira &  OH/IR  &  1.44$\pm$0.28&         7 &  1.01$\pm$0.28&        11 &  1.11$\pm$0.22&        12 &  0.96$\pm$0.09&         4 &  1.19$\pm$0.09&         3 \\
      Mira & oxygen  &  1.44$\pm$0.44&        32 &  1.08$\pm$0.28&        39 &  1.20$\pm$0.27&        35 &  0.83$\pm$0.15&         4 &  1.18$\pm$0.03&         2 \\
      Mira & carbon  &  2.76     &         1 &  0.88$\pm$0.25&         6 &  1.16$\pm$0.20&         7 &  1.06$\pm$0.42&         8 &  1.14$\pm$0.06&         2 \\
      Mira &      S  &  1.66     &         1 &  0.99$\pm$0.01&         2 &  1.37$\pm$0.36&         2 &       $-$    &         0 &       $-$    &         0 \\
       SRa & oxygen  &  0.95$\pm$0.51&         4 &  1.19$\pm$0.18&         5 &       $-$    &         0 &       $-$    &         0 &       $-$    &         0 \\
       SRa & carbon  &  2.06$\pm$0.15&         2 &  1.29     &         1 &       $-$    &         0 &  1.31     &         1 &       $-$    &         0 \\
       SRb & oxygen  &  1.42$\pm$0.21&         6 &  0.81$\pm$0.19&        10 &  1.47$\pm$0.35&         5 &       $-$    &         0 &       $-$    & 0 \\
       SRb & carbon  &  2.28$\pm$0.80&         3 &  1.14     &         1 &  0.97     &         1 &       $-$    &         0 &       $-$    &         0 \\
\enddata
\end{deluxetable}

\end{document}